\documentclass[aps,pra,reprint]{revtex4-1}

\usepackage{amssymb,amsmath,latexsym,amscd,amsfonts}  
\usepackage{graphicx}  
\usepackage[usenames]{color}
\usepackage{umoline}

\usepackage{graphicx}% Include figure files
\usepackage{dcolumn}% Align table columns on decimal point
\usepackage{bm}% bold math

\newtheorem{theorem}{Theorem}
\newtheorem{definition}{Definition}
\newtheorem{claim}[theorem]{Claim}

\newtheorem{lemma}[theorem]{Lemma}

\newcommand{\qed}{\mbox{\ \ \ }\rule{6pt}{7pt} \medskip}
\newenvironment{proof}{\noindent{\em Proof:}}{\hfill\qed}

\newcommand{\ket}[1]{|#1\rangle}
\newcommand{\bra}[1]{\langle#1|}

\newcommand{\braket}[2]{\langle {#1} | {#2} \rangle}

\newcommand{\Eq}[1]{Eq.~(\ref{#1})}
\newcommand{\Fig}[1]{Fig.~\ref{#1}}
\newcommand{\Lem}[1]{Lemma~\ref{#1}}
\newcommand{\Thm}[1]{Theorem~\ref{#1}}
\newcommand{\Sec}[1]{Sec.~\ref{#1}}

\newcommand{\Def}[1]{Definition~\ref{#1}}
\newcommand{\EqDef}{\stackrel{\mathrm{def}}{=}}
\newcommand{\Ref}[1]{Ref.~\cite{#1}}

\newcommand{\BG}{\bm{\Gamma}}
\newcommand{\BB}{\bm{B}}
\newcommand{\BC}{\bm{C}}
\newcommand{\BA}{\bm{A}}
\newcommand{\BDel}{\bm{\Delta}}

\newcommand{\Bl}{\bm{\lambda}}
\newcommand{\Bm}{\bm{\mu}}
\newcommand{\bigO}{\mathcal{O}}

\newcommand{\Ttri}[1]{{\Gamma^{[#1]}}}
\newcommand{\Td}[1]{\lambda^{[#1]}}

\newcommand{\TR}[1]{\mathbf{R}^{[#1]}}
\newcommand{\TL}[1]{\mathbf{L}^{[#1]}}
\newcommand{\norm}[1]{{\| #1 \|}}

\newcommand{\calh}{{\cal{H}}}

\begin{document}

\title{An Efficient Algorithm for approximating 1D Ground States}

\author{Dorit Aharonov}
  \email{dorit.aharonov@gmail.com}
  \affiliation{ School of Computer Science and Engineering, \\
  Hebrew University, Jerusalem, Israel}

\author{Itai Arad}
  \email{arad.itai@gmail.com}
  \affiliation{ School of Computer Science, \\
  Tel-Aviv University, Tel-Aviv, Israel}
  
\author{Sandy Irani}
  \email{irani@ics.uci.edu.}
    \affiliation{Computer Science Department, \\
  University of California, Irvine , CA, USA}
\date{\today}

\begin{abstract}
  The density-matrix renormalization-group method is very effective
  at finding ground states of one-dimensional (1D) quantum systems
  in practice, but it is a heuristic method, and there is no known
  proof for when it works.  In this article we describe an efficient
  classical algorithm which provably finds a good approximation of
  the ground state of 1D systems under well defined conditions. More
  precisely, our algorithm finds a matrix product state of bond
  dimension $D$ whose energy approximates the minimal energy such
  states can achieve. The running time is exponential in $D$, and so
  the algorithm can be considered tractable even for $D$ which is
  logarithmic in the size of the chain. The result also implies
  trivially that the ground state of any local commuting Hamiltonian
  in 1D can be approximated efficiently; we improve this to an exact
  algorithm.
\end{abstract}
\maketitle


%
% ======================================================================
%

\section{Introduction}

Finding ground states of local one-dimensional (1D) Hamiltonian
systems is a major problem in physics. The most commonly used method
is the density-matrix renormalization-group (DMRG)
\cite{White92,White93,ostlund1, ostlund2, peschel, DMRG-overview},
discovered in 1992. DMRG can be cast in the form of matrix product
states (MPSs) which are succinct representations of 1D quantum states
using $D\times D$ matrices, where the coefficients in the state can
be written in terms of products of these matrices.  The number of
matrices is $dn$, where $d$ is the dimension of each individual
particle and $n$ is the number of particles in the system. The
parameter $D$ is called the \emph{bond dimension}.  DMRG works
essentially as follows: The algorithm starts with some initial MPS
and sweeps from one end of the chain to the other, optimizing the
entries of the matrices at one site with the other parameters fixed.
Some versions allow optimizing over two neighboring sites at once,
which enables the algorithm to increase the bond dimension in the
course of the algorithm for improved accuracy. In all cases, the
approach is to apply local optimizations iteratively.  It is thus
easy to construct examples in which the DMRG algorithm gets trapped
in a local minimum. To illustrate this, think of a 1D
spin chain whose Hamiltonian consists of two types of interactions:
One type consists of interactions which force the spins to be
aligned; every two neighboring sites gain an energy penalty of say
$4$ if they are not aligned. The other type of term gives every spin
an energy penalty of $1$ if it points upward. Starting from the
all-up string, a local move only increases the energy; thus, local
update rules cannot take the system to its ground state, the
all-down string.  This example can of course be handled by
randomizing the initial string, for example, or increasing the
window size; however, it demonstrates that DMRG has a fundamental
difficulty in addressing non local characteristics of the system. It
is natural to ask if there is a general algorithm that does not get
stuck in local minima as DMRG does and provably always find a good
approximation of the ground state of a given 1D system in a
reasonable amount of time.

To answer this question, we first ask what is known regarding the
analogous question in the easier, classical, case.  It was Kitaev
\cite{kitaev:book} who drew the important connection between the
problem of finding ground energy and ground states of local
Hamiltonians, and the well-known classical \emph{constraint
satisfaction problem} (CSP). The input to a CSP consists of
constraints $\{H_c\}_c$ on $n$ $q$-state classical particles. Each
$H_c$ acts on $k$ particles (for some constant $k$) and is given as
a Boolean function on the possible assignments to those $k$
particles; when $H_c=1$ the configuration is forbidden and when
$H_c=0$ it is allowed. The problem is to determine the maximum
number of constraints that can be satisfied, or alternatively, to
minimize $\sum_c H_c$.  The decision version of this problem is to
determine whether it is possible to satisfy more than some given
number of constraints. This is one of the most well-known
NP-complete problems.  CSP can clearly be seen as a special case of
the problem of finding ground states and ground energies of local
Hamiltonians, in which the terms in the Hamiltonian are projections
on local forbidden configurations.  This analogy has led over the
past few years to many interesting insights regarding the local
Hamiltonian problem (see, e.g., \Ref{kitaev:book, bravyi,
adiabatic, focsversion, detectability, randomsat}).

Let us therefore see what the known classical results regarding CSP
in 1D can teach us about 1D local Hamiltonians and their ground
states.  We recall that in the classical case, 1D CSPs (in which the
particles are arranged in a line and constraints are between $k$
adjacent neighbors) are dramatically easier than their
higher-dimensional counterparts. While even the 2D case is
NP complete, the 1D problem can be solved in polynomial time. The
reason for the tractability of the problem in 1D is essentially that
the problem can be divided into sub problems, namely, the left- and
the right-hand sides of the chain, which interact only via the $k$
particles on the border. The fact that these particles can only be
assigned a small number of possible values makes it possible to
handle the problem by solving each sub problem separately for each
fixed possible assignment to the border particles and then gluing
the sub solutions together by picking the best choice for the middle
particles.  We explain the algorithm in detail later; the
outcome is an algorithm which is linear in the number of particles
in the chain and quadratic in the number of states per particle. 
      
Unfortunately, there is no hope of getting such a general result for
the 1D quantum problem.  Aharonov \emph{et al} \cite{focsversion}
have shown that approximating the ground energy for general 1D
quantum systems is as hard as quantum-NP.  Even when restricted to
ground states that are well-approximated by MPSs of polynomial bond
dimension, the problem is NP-hard, as was shown by Schuch \emph{et
al} \cite{schuch08}. A related earlier result due to Eisert
\cite{eisert06} showed that optimizing a constant number of matrices
in the MPS representation subject to fixed values in the other
matrices is NP-hard.  These results indicate that the dichotomy
between the computational difficulty of 1D and 2D classical systems
does not carry over to the quantum setting, and it is highly
unlikely that the quantum 1D problem is tractable. Nevertheless, we
show here that using the classical 1D algorithm as a template for an
algorithm for the quantum problem leads to a solution for a wide
and interesting class of local Hamiltonian problems, namely, for
those cases in which we can assume that the bond dimension is small.

%
%-----------------------------------------------------------------------------
%

\subsection{Main Result}
\label{sec:results}  

We derive an efficient algorithm for approximating the minimal
energy of a 1D system among all states of a bounded bond dimension
$D$. The algorithm is exponential in $D$ and thus can be considered
reasonable, though maybe not practical, even for $D$, which is
logarithmic in the size of the chain.  The algorithm also provides a
description of an MPS with the approximate minimal energy. 

\begin{theorem}
\label{thm:mps} Let $H$ be a nearest-neighbor Hamiltonian on a 1D
  system of $n$ $d$-dimensional particles. Let $J$ be a bound on the
  operator norm of each local term. There is an algorithm
  that takes as input $\epsilon$, $H$ and $D$ and produces an MPS
  $\ket{\Omega}$ of bond dimension $D$, such that for any MPS
  $\ket{\psi}$ of bond dimension $D$ with $nD^2\ge 12$, 
  \begin{align}
  \label{eq:mps}
    \bra{\Omega}H\ket{\Omega} \le
      \bra{\psi}H\ket{\psi}+2JD^2 n^2\epsilon \ . 
  \end{align}
  The algorithm runs in time $n \cdot poly(d,D,N)$, where $N =
  \mathcal{O}\left(\frac{144dD}{\epsilon}\right)^{D + 2dD^2}.$
\end{theorem}
Several remarks are in place here. First, note that the restriction
that the interactions are nearest neighbor is done without loss of
generality since any 1D system can be reduced to a 2-local 1D
system with nearest neighbor interactions by grouping neighboring
particles together. 

Note also that the running time in the above theorem is phrased as a
linear function in $n$, the size of the system, times some fixed
amount of time spent per particle. The error, however, scales with
$n^2$.  One may want to apply the theorem to derive an approximation
with a fixed additive error $\delta$, in which case simply set
$\epsilon =\delta n^{-2}$ in the above theorem to get the running 
time as a function of $\delta$. 

This result shows that the problem of finding bounded bond dimension
MPSs can be done in polynomial time.  Unfortunately, the running
time, though efficient in theory, is quite impractical, as even for
$D=2$ and the error $\epsilon/n^2$ a constant, we get a running
time which scales like $n^{16}$.  It is hard to imagine that these
running times are practical. Nevertheless, it is very likely that
the running time can be improved; in particular, when solving
specific problems with certain symmetries, dramatic improvements may
be possible.  Moreover, it is possible that this algorithm can be
used to boost DMRG in certain cases where it gets stuck or to
create the initial state of DMRG.  All these improvements are left
for further research.

We now provide an overview of the algorithm.  To understand the
general idea, we first recall how the classical 1D algorithm works
in detail.  Consider the case of the classical CSP on a line with
$k=2$, namely the problem of minimizing the energy function
$H=\sum_{i=1}^n H_{i,i+1}$.  An optimal assignment can be found
efficiently by a standard algorithmic technique called \emph{dynamic
programming}.  Define the partial problem up to the $(r+1)$th
particle, $H_r=\sum_{i=1}^r H_{i,i+1}$. The algorithm starts with
the partial problem defined for $r=1$ and creates a list $L_2$ of
possible assignments to the first two particles as follows: For
each of the $q$ possible assignments $\sigma_2$ to particle $2$, the
algorithm finds an assignment $\sigma_1$ to particle $1$ which
minimizes $H_1(\sigma_1,\sigma_2)$.  That optimal $\sigma_1$ is
called the {\it tail} of $\sigma_2$.  For each $\sigma_2$ the
algorithm keeps its tail $\sigma_1$ and also the energy of this
partial assignment, $H_1(\sigma_1,\sigma_2)$. $L_2$ thus contains 
the best possible partial assignment with each possible ending.
After $r-1$ iterations, we assume the algorithm has a list $L_r$
consisting of an optimal tail $\sigma_1,...,\sigma_{r-1}$ for each
of the $q$ possible assignments $\sigma_r$ to the $r$th particle,
where optimality is measured with respect to $H_{r-1}$. In other
words, the algorithm has a solution to the subproblem confined to
the first $r$ particles, with any possible ending.  To include the
next particle, and create the next list $L_{r+1}$, the algorithm
finds the optimal tail of each assignment $\sigma_{r+1}$. This is
done by considering all items in the list $L_r$ as possible tails
for $\sigma_{r+1}$ and taking the tail which minimizes
$H_{r}(\sigma_1,...,\sigma_{r+1}).$ In each of the $n-1$ iterations,
the algorithm checks for each of the $q$ possible assignments
$\sigma_r$, all $q$ items in the list $L_{r-1}$. Thus, in time which
is linear in $n$ and quadratic in $q$, we can derive the final list
$L_{n-1}$. The final solution is an assignment of minimal energy in
that list. 

The main idea in this article is to generalize the above algorithm
to MPSs by replacing assignments to particles by possible values of
MPS matrices. Since matrices are continuous objects, we use an
$\epsilon$-net over all possible matrices of bond dimension $D$. The
number of possible assignments to one variable, $q$, will now be
replaced by the number of points in the $\epsilon$-net, denoted as
$N$.  We will move from one site to the next, keeping track of the
minimum-energy MPS state, which ends in each MPS matrix for the
right most particle that the algorithm has reached. 

In order to carry out this idea, it must not happen that the choice
of the MPS matrix of a later iteration can change the optimality of
the partial MPS state found in an earlier iteration.  To avoid this,
we work with a restricted form of MPSs called \emph{canonical 
MPSs}, in which the energy of each
term in the Hamiltonian depends only on MPS matrices associated with
nearby particles. There are, however, various technical issues we
need to handle. In particular, we cannot use perfectly canonical
MPSs but only an approximated version of those, which imposes
further technicalities, and in particular, the 
neighboring MPS matrices do not match perfectly (we call this 
\emph{imperfect stitching}).  
These technicalities make the error analysis a bit subtle.
Before we formally define canonical MPSs and provide the details of
the algorithm, we mention an implication for a related problem. 

\subsection{Commuting Hamiltonians in 1D}
\label{sec:commutingIntro} 

A problem related to finding minimum-energy MPS states is the
complexity of calculating the ground energy of commuting
Hamiltonians in which all the local terms commute. 
Bravyi and Vyalyi proved that for 2-local Hamiltonians
the problem lies inside NP \cite{bravyi}.  For $k$-local commuting
Hamiltonians with $k>2$, the complexity of the problem is still
open. The complexity of the 1D case was not studied before as far as
we know; an immediate corollary of \Thm{thm:mps} is that
there is an efficient classical algorithm for approximating the
ground energy of commuting Hamiltonians in 1D to within $1/poly(n)$.
This is because the ground state of a commuting Hamiltonian in 1D is
an MPS of constant $D$ (this is a well-known fact that we explain
later for completeness), and therefore \Thm{thm:mps} can be
applied. In fact, the result can be improved to an exact algorithm
(up to exponentially good approximations due to truncations of real numbers) 
for a certain general class of problems. We prove the following.
\begin{theorem}
\label{thm:comm} Given is a 1D Hamiltonian whose terms commute.
  There is an efficient algorithm that can compute the ground energy
  of this Hamiltonian to within any desired accuracy $\epsilon$ in
  time polynomial in $n$ and in $\frac{1}{\epsilon}$. If we may
  assume also that the ground space of the total Hamiltonian is well
  separated from the higher excited states, by a spectral gap which
  is at least $1/poly(n)$, then the algorithm can find both the
  ground energy and a description of an MPS for the ground state
  exactly (i.e., up to exponentially small errors due to handling of real
  numbers). 
\end{theorem}

The basic idea for the exact algorithm can be illustrated when the
terms in the Hamiltonian are all projections and the ground state 
is unique. Since the terms commute, the ground state is an
eigenstate of each term separately, with eigenvalue either $0$ or
$1$.  We start by applying the dynamic programming algorithm, to
create a good approximation of the ground state.  From this
approximation we can deduce the correct eigenvalue ($0$ or $1$) for
each of the terms. The projections on the relevant eigenspaces can
then be applied to the MPS of the approximate state to make it
exact.  One gets a tensor network of small depth, which can be
converted into an MPS again. It can be shown that applying the
projections does not increase the bond dimension of the MPS too much
with respect to the approximating state. The details are fleshed
out in the proof (\Sec{sec:commuting}). 

Handling the degenerate case is very easy; essentially, we force
the dynamic algorithm to choose one state of the various possible
states. The assumption on the spectral gap ensures that the
errors created by the epsilon net approximations would not cause a
confusion between the ground space and some excited states. 

We provide an alternative proof of \Thm{thm:comm}, which also uses
dynamic programming. In fact, this proof holds for a somewhat
stronger version of the theorem, in which the conditions on the
spectrum are far less restrictive.  In the algorithm given by this
approach, the state is not provided as an MPS but rather as a tensor
product of two-particle states. The construction is based on the
work of \Ref{bravyi} in which it is proved that the ground states of
$2$-local commuting Hamiltonians have this special structure. Bravyi
and Vyalyi use this structure to show that general $2$-local
commuting Hamiltonians problem is in NP.  Since 1D chains with
$k$-local interactions can always be made $2$-local by treating
nearby particles as one particle of a larger dimension, \Ref{bravyi}
implies that the 1D commuting problem lies in NP. However, by
exploiting the special form of these ground states, dynamic
programming can be applied to find the solution efficiently in a
very similar manner to the 1D CSP, in which the NP witness is found
using the 1D structure. Unfortunately, in this approach too, it
seems that one cannot avoid some assumption on the spectrum of the
total Hamiltonian, albeit a significantly less restrictive one.  
Throughout its execution, the dynamic programming algorithm compares
various partial energies. If these are too close, and cannot be
distinguished even by computations performed with exponentially good
precision, then the algorithm might get confused between the ground
energy and a slightly excited state.  A sketch of the alternative
proof of \Thm{thm:comm}, providing the stronger version of it, and a
discussion of the above precision issue are given in
\Sec{sec:commuting}.

We mention that this latter proof (and in particular the observation
that dynamic programming can be useful for 1D quantum systems and
not only for 1D classical systems) was the inspiration for the
current article, rather than its corollary. 

%
%------------------------------------------------------------------------
%
\subsection{Discussion and Open Questions}
\label{sec:conc} 

It is natural to ask how much the results in this article can be
improved.  By \Ref{schuch08}, we know that no polynomial algorithm
exists for finding optimal approximations of \emph{polynomial} bond
dimension (unless P=NP).  However, the difficult instances of
\Ref{schuch08} have a spectral gap of $1/poly(n)$.  Hastings has
shown that ground state of 1D quantum system with a constant gap
can be approximated by a MPS with polynomial bond dimension
\cite{hastings}. However, this is too large to immediately yield an
efficient algorithm from our result. It may still be true, however,
that under the additional restriction that the Hamiltonian has a
constant gap, a polynomial time algorithm exists, even when the bond
dimension is as large as polynomial. 

It is very likely that the efficiency of our algorithm can be
significantly improved even for the general case.  In particular, a
factor of $n$ would be shaved from the error in \Thm{thm:mps}
if we could use an $\epsilon$-net which is both exactly canonical
and enables perfect overlap between matrices at neighboring
particles, as we later explain. Unfortunately, even if this can be
done, the running time for this general algorithm is still quite
large. 

As mentioned earlier, we leave for further research the question of 
how this algorithm can be used in combination with DMRG, and how
certain symmetries in the problem can be utilized to enhance its
performance time for specific interesting cases.  

We note that very similar results to those presented in this 
article were derived independently by Schuch and Cirac \cite{NI}. 

%-------------------------------------------------------------
\subsection{Paper Organization}

{~}

Section~\ref{sec:TandM} starts by defining tensor networks, MPSs and
canonical MPSs.  In \Sec{sec:Algorithm} we describe the algorithm.
This is where the $\epsilon$-nets are defined and an algorithm to
generate them is given. Also in \Sec{sec:Algorithm}, we show how
they are used in the dynamic programming algorithm.  \Sec{sec:Error}
provides an exact analysis of the error accumulated in the
algorithm. The complexity is analyzed as a function of the desired
error.  In \Sec{sec:commuting} we provide the proof regarding the
approximate and exact solutions for the commuting 1D case.  We defer
several technical lemmas to the Appendix.

%
% ======================================================================
%

\section{Tensor Networks and Matrix Product States}
\label{sec:TandM}

% --------------------------------------------------------------------
\subsection{Tensor Networks}
\label{sec:TensorNetworks}

We start with some background on tensor networks, since MPSs are a
special case of those. A detailed introduction to the use of
tensor networks in the context of quantum computation can
be found in Refs~\cite{markov2008simulating, aharonov2006quantum,
arad2008quantum}.

A tensor network is a graph in which we allow some of the edges to
be incident to only one node. These edges are called the \emph{legs}
of the network. Each node is assigned a tensor whose rank (number of
indices) is equal to the degree of the node. Each index of the
tensor corresponds to one edge that is incident to that node.  To
each edge (or index) we also assign a positive integer which
indicates the range of the index.  The indices associated with some of
the edges in the tensor network may be assigned fixed values. The
other edges are called \emph{free} edges.

We call an assignment of values to the indices of the free edges in
the network a \emph{configuration}. With all the indices fixed, the
tensor at each node in the network yields a particular value. We say
that the value of the configuration is the product of the values for
each of the nodes.

The value of the network is in general a tensor, whose rank is equal
to the number of legs in the network. If there are no such legs, the
value is simply a number (a scalar).  Each assignment of values to
the indices associated with the legs of the network gives rise to a
value for the network tensor. We compute the tensor value for this
assignment by summing over all configurations which are consistent with that
assignment the value of each such configuration.

We note that often in the literature, one assigns values not to
entire edges but to the two sides of an edge separately (where each
side inherits its range of indices from the tensor associated with
the node on that side).  In the evaluation of the network, we
require that the values on the two sides of one edge are equal, or
else the entire configuration contributes zero to the sum. 

Tensors will be denoted as bold-face fonts: $\Bl, \BG, \Bm$. Their
contraction will be denoted as an expression like $\Bl\BG\Bm$, when
it is clear from the context along which indices the contraction is
performed.

It is possible to restrict a tensor of rank $k$ to a tensor of 
rank $k-1$ by assigning a fixed value to one of its legs. 
For example, $\BG_\alpha$ is the restriction of the tensor $\BG$ 
to the case in which the relevant edge associated with the index $\alpha$ 
is given some value (which, by the usual abuse of notation of
  variables and their values, will also be denoted as $\alpha$).

It is convenient to associate with every tensor (which can be given
as a contraction of a tensor network) a quantum state. For example,
let $\BG=\Gamma^i_{\alpha,\beta}$ be a rank-$3$ tensor. Then we
define $\ket{\BG}\EqDef \sum_{i,\alpha,\beta}
\Gamma^i_{\alpha,\beta} \ket{\alpha}\otimes\ket{i}\otimes\ket{\beta}$.

%
%-----------------------------------------------------------------------------
\subsection{Matrix Product States} \label{sec:mps}

\begin{figure*}
  \begin{center}
    \includegraphics[scale=1]{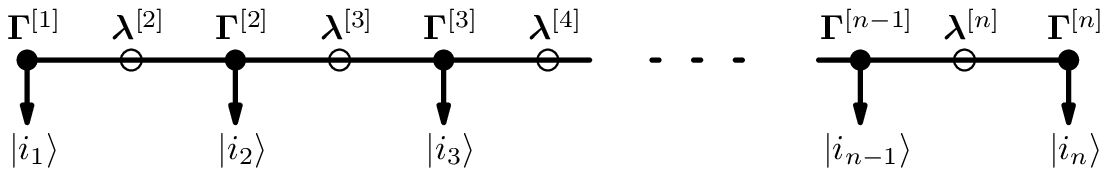}
  \end{center}
  \caption{MPS as a tensor network. \label{fig:mps} }
\end{figure*}

We work in the notation of Vidal \cite{vidal1} for MPSs, with minor
changes.  A MPS of a chain of $n$ $d$ dimensional particles, with
bond dimension $D$, is a tensor network with a 1D structure as in
\Fig{fig:mps}.  Horizontal edges correspond to indices ranging from
$1$ to the bond dimension $D$ and are denoted with $\alpha,
\beta$,..., while vertical edges correspond to indices ranging from
$1$ to the physical dimension $d$. (In our description, the end
particles will actually have a different physical dimension, denoted
$d_{end}$. This is required due to a technical reason described in
\Sec{sec:canonical}) The indices of vertical edges are denoted with
$i$,$j$,\ldots The figures show two types of nodes: black and white.
The tensors of black nodes are typically of rank $3$ (except for the
boundary tensors, which are of rank $2$), and we denote them with
$\BG$'s. For example, when the tensor that is second from left is
written with its indices, it is denoted as
$\Gamma^{{[2]}^i}_{\alpha_2, \alpha_3}$, where the index $[2]$ in
brackets corresponds to its location in the graph.  The tensors
associated with white nodes are always of rank $2$ and are denoted
with $\Bl$'s. They are required to be diagonal and hence are given
only one index (i.e., $\lambda^{[2]}_{\alpha_2}$). Without loss of
generality, we will also demand that the entries of $\Bl$ are
non negative since the phases can be absorbed in the neighboring
$\BG$ tensors.

The MPS defined by this network is
$\ket{\psi} = \sum_{i_1, \ldots, i_n} C_{i_1\cdots i_n}
\ket{i_1}\cdots\ket{i_n}$ with 
\begin{align*}
  C_{i_1\cdots i_n} \EqDef  
  \sum_{\alpha_2, \ldots, \alpha_n}
    \Ttri{1}_{\alpha_2}^{i_1} \Td{2}_{\alpha_2}
        \Ttri{2}_{\alpha_2\alpha_{3}}^{i_2}\Td{3}_{\alpha_3}\cdots
    \Td{n}_{\alpha_n}\Ttri{n}_{\alpha_n}^{i_n} \ .
\end{align*}

\begin{figure*}
  \begin{center}
    \includegraphics[scale=1]{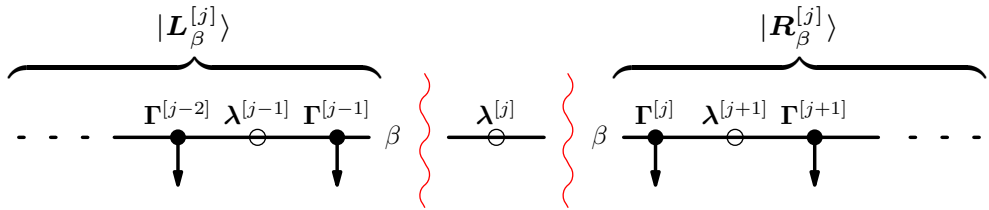}
  \end{center}
  \caption{A description of a canonical MPS. The tensors are chosen
  such that cutting a MPS between the $j-1$th and $j$th
  particles corresponds to the Schmidt decomposition between the left and
  right parts: $\ket{\psi} = \sum_\beta
  \lambda^{[j]}_\beta\ket{\TL{j}_\beta}\otimes\ket{\TR{j}_\beta}$.
  \label{fig:cut} }
\end{figure*}

In the language of tensor states, $\ket{\psi}$ is exactly 
the tensor state of the contraction 
$\BG^{[1]}\Bl^{[2]}\BG^{[2]}\cdots\Bl^{[n]}\BG^{[n]}$.

%
%-----------------------------------------------------------------
%
\subsection{Canonical MPSs}
\label{sec:canonical}

An MPS is in \emph{canonical form} if every cut in the chain induces
a Schmidt decomposition (as in \Fig{fig:cut}). In other words, we
can rewrite the MPS by changing the order of summation to sum last
over the index $\beta$ of the $j$th $\Bl$ tensor: $ \ket{\psi} =
\sum_{\beta} \Td{j}_{\beta} \ket{\TL{j}_\beta} \otimes
\ket{\TR{j}_\beta}$, where $\TL{j}_\beta$ ($\TR{j}_\beta$) denote
the contraction of the all the tensors to the left (right) of the
cut with fixed $\beta$ and $\ket{\TL{j}_\beta}$ ($\ket{\TR{j}_\beta}$) are
their corresponding states. Then the canonical conditions are that
for all $j$ from $2$ to $n$, $\sum_\beta \big(\Td{j}_\beta\big)^2 =
1$ and $\bra{\TL{j}_\alpha}\TL{j}_\beta\rangle =
\bra{\TR{j}_\alpha}\TR{j}_\beta\rangle = \delta_{\alpha\beta}$.  In
addition, for normalization, we require that the entire MPS state is
normalized, which is guaranteed by the normalization requirement on
the $\Bl^{[j]}$ tensors.

There is a small technical issue that needs attention: The canonical
conditions cannot be satisfied at the boundaries if $d<D$.
Consider for example the left boundary; there are not enough
dimensions in the Hilbert space of the left particle for an
orthonormal set of vectors $\ket{\TL{2}_\alpha}$ to exist.  This
issue remains a problem even as we move away from the boundary by
one particle, as the dimension of the left-side Hilbert space
increases to $d^2$ which may still be smaller than $D$.  There are
many ways of handling this technicality; here we choose to assume that
the particles at the end of the chain have dimension of at least $D$.
This will ensure that at any cut along the chain, the Hilbert space
of the subsystems on each side have dimension of at least $D$.  We can
achieve this by grouping $s$ particles at each end of the chain into
a single particle, where $s$ is chosen to be the smallest integer
such that $d^s \ge D$. Denote $d^s$ as $d_{end}$, the dimensionality
of each of those end particles.  Note that $d_{end} = d^s \le Dd$.
The dimension of the rest of the particles will remain $d$. We
renumber the particles after the grouping, so that the new $H_{1,2}$
is now the sum of the old $H_{i,i+1}$ for $i$ ranging from $1$ to
$s$. The term in the Hamiltonian for the last two particles is
adjusted in a similar manner. We will assume from now on that the
Hamiltonian is given in this form.  

Let us now see how the canonical conditions can be stated in a local
manner. Graphically, the second condition is equivalent to
\begin{align}
\label{eq:c-cond}
  \includegraphics[scale=0.6]{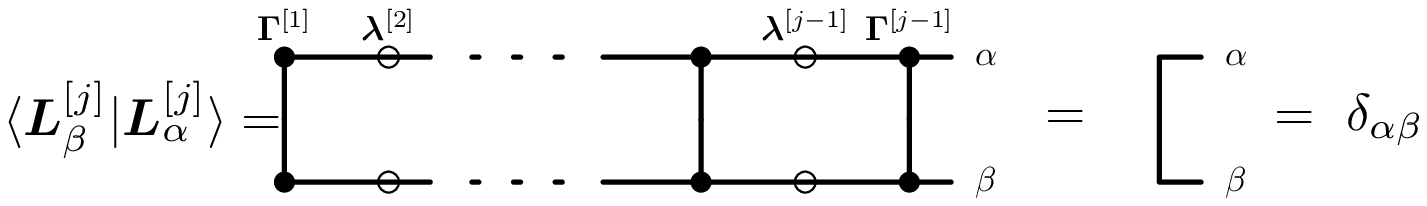} 
\end{align}
and similarly from the other side. Here the upper part of the
network corresponds to $\ket{\TL{j}_\alpha}$, and the lower part
corresponds to $\bra{\TL{j}_\beta}$.  Notice that the canonical
conditions imply that we can ``collapse'' the network both from the
left side and from the right side. Moreover, as this condition holds
at every bond, it is not difficult to see that a necessary and
sufficient condition for an MPS to be canonical consists of the following
\emph{local} conditions on {$(\Bl^{[j]},\BG^{[j]},\Bl^{[j+1]})$}:
For every {$j=2, \ldots, n-1$}, 
\begin{align}
  \langle (\Bl^{[j]} \BG^{[j]})_\alpha|(\Bl^{[j]}\BG^{[j]})_\beta\rangle 
    &= \delta_{\alpha\beta} 
      \ \ \text{\small (left canonical)},
     \label{eq:cond-left}\\
  \langle (\BG^{[j]} \Bl^{[j+1]})_\alpha|(\BG^{[j]}\Bl^{[j+1]})_\beta\rangle 
    &= \delta_{\alpha\beta}
     \ \ \text{\small (right canonical)}. \label{eq:cond-right}
\end{align}

\noindent
For $j=1$ and $j=n$, for $1\le\alpha,\beta\le D$:
\begin{align} \label{eq:cond-bound}
    &\langle \BG^{[1]}_\alpha| \BG^{[1]}_\beta\rangle =
    \langle \BG^{[n]}_\alpha | \BG^{[n]}_\beta\rangle
    = \delta_{\alpha\beta} \\
    &\qquad \text{(boundary canonical conditions)}  \ . \nonumber
\end{align} 

\begin{figure*}
  \begin{center}
    \includegraphics[scale=0.9]{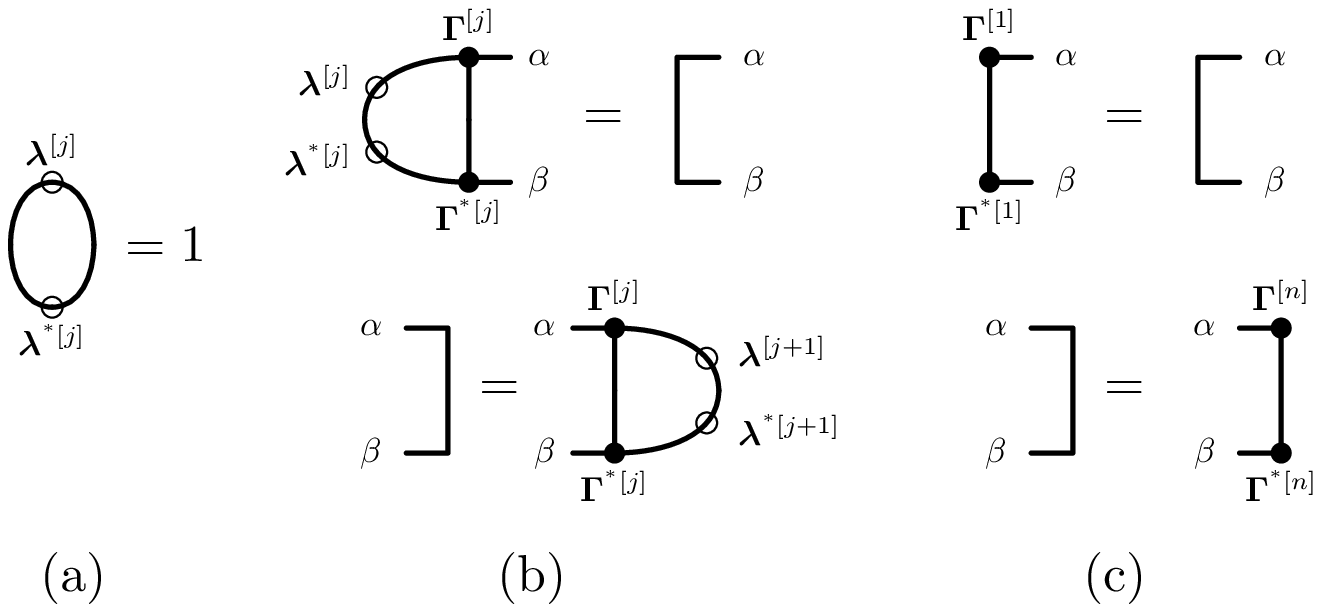}
  \end{center}
  \caption{ \label{fig:cond}
    (a) The normalization condition for $j=2, \ldots, n$.
    (b) The left/right canonical conditions  for $j=2, \ldots, n-1$
    [see Eqs.~(\ref{eq:cond-left}) and (\ref{eq:cond-right})].
    (c) The boundary canonical conditions for $j=1$ and $j=n$ [see
    \Eq{eq:cond-bound}]. }
\end{figure*}

We also require that the $\Bl's$ are normalized, namely, 
that for every $j$ from $2$ to $n$, 
\begin{align} \label{eq:cond-norm} 
% \text{\textbf{Normalization:}} \qquad
  \langle \Bl^{[j]} |\Bl^{[j]} \rangle = 1 \ .
\end{align} 
Graphically, these conditions are summarized in \Fig{fig:cond}.

Any triplet $(\Bl^{[j]}, \BG^{[j]}, \Bl^{[j+1]})= (\Bl, \BG, \Bm)$
that satisfies the normalization and the left and right canonical
conditions [Eqs.~(\ref{eq:cond-left}), (\ref{eq:cond-right}),
and (\ref{eq:cond-norm})] is called a \emph{canonical triplet}. Such a
triplet can be associated with a quantum state on three particles
$\ket{\psi} = \ket{\Bl\BG\Bm}= \sum_{\alpha, i, \beta}
\lambda_\alpha \Gamma^i_{\alpha\beta} \mu_\beta
\ket{\alpha}\ket{i}\ket{\beta}$, with the following properties:
$\norm{\psi} = 1$; the Schmidt basis of the first particle is the
standard basis, with Schmidt coefficients $\{\lambda_\alpha\}$; and
the Schmidt basis of the third particle is the standard basis, with
Schmidt coefficients $\{\mu_\beta\}$. A canonical MPS can thus be
described as a set of canonical triplets (or equivalently
$3$-particle states) such that the right $\Bm$ tensor of one state
is equal to the left $\Bl$ tensor of the next canonical triplet.

Instead of describing a canonical MPS in terms of canonical triplets
$(\Bl, \BG, \Bm)$, we will often describe it using \emph{canonical
pairs} $(\Bl, \BB)$, where 
\begin{align*}
  \BB \EqDef \BG\Bm \ .
\end{align*}
The advantage is that for canonical MPSs, the elements in $\BB$ are
always bounded (since the $L_2$ norm of $\BB$ satisfies
$\norm{\BB}=\sqrt{D}$; see \Sec{sec:norms}), unlike $\BG$ whose
entries can approach infinity when the corresponding $\Bm$ entries
approach zero.

An MPS that is described by the contraction
$\BG^{[1]}\Bl^{[2]}\BG^{[2]}\Bl^{[3]}\cdots \Bl^{[n]}\BG^{[n]}$ can
also be denoted as $\BG^{[1]}\Bl^{[2]}\BB^{[2]}\BB^{[3]}\cdots
\BB^{[n-1]}\BG^{[n]}$.  No information is lost since $\Bm$ can
always be recovered from $(\Bl,\BB)$: $\mu_{\beta}$ is the norm (see
\Sec{sec:norms}) of the tensor state
$(\Bl\BB)_{\beta}$:\footnote{Recall that $\mu_{\beta}$ corresponds
to a Schmidt coefficient in a Schmidt decomposition that coincides
with the standard basis.}
\begin{align*}
  \mu_\beta = \left(\sum_{i,\alpha}|\lambda_\alpha
    B_{\alpha\beta}^i|^2\right)^{1/2} \ .
\end{align*}
We define $\Bm\EqDef\Bm(\Bl,\BB)$
this way also for non-canonical pairs. 

The advantage of working with the canonical form is that the energy
of local Hamiltonians involves only the local tensors, as the
following figure illustrates:
\begin{center}
  \includegraphics[scale=0.53]{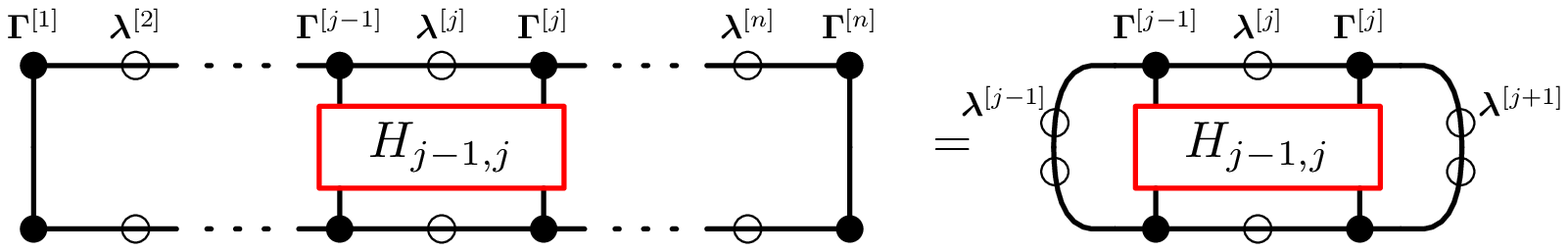} 
\end{center}
The above equality was obtained using the canonical conditions that
are described in \Eq{eq:c-cond}. Consequently, the energy
$\bra{\psi}H_{j-1,j}\ket{\psi}$ only involves five tensors:
$\Bl^{[j-1]}, \BG^{[j-1]}$, $\Bl^{[j]}, \BG^{[j]}$, and
$\Bl^{[j+1]}$. Similarly, $H_{1,2}$ only depends on $\BG^{[1]}$,
$\Bl^{[2]}$, $\BG^{[2]}, \Bl^{[3]}$, and $H_{n-1,n}$ only depends on
$\Bl^{[n-1]}, \BG^{[n-1]}$, $\Bl^{[n]}, \BG^{[n]}$.  It is important
that each energy term does not involve tensors further to the right
in the chain since the algorithm attempts to compute (or
approximate) the optimal MPS up to a certain point. We would like to
be able to grow the description of the state from left to right,
without affecting the energies we have already computed. If matrices
in the right side of the chain affected energies of terms in the
left side, we would need to go back and change the MPS matrices of
the particles we have already handled after we make new assignments
to particles on the right.  This would ruin the entire idea of
dynamic programming. 

Fortunately, any MPS representing a normalized state can be written
as a canonical MPS with no increase in bond dimension. This follows
from \Ref{vidal1}, in which it is shown that any state with Schmidt
rank of at most $D$ across any cut can be written as a canonical MPS
with bond dimension $D$.

%-------------------------------------------------------------------------------
\subsection{Tensor Norms and Distances}\label{sec:norms} 

We use the $L_2$ norm on tensors $\norm{\bm{X}}^2 \EqDef
\sum_{i_1\ldots i_k} |X_{i_1\ldots i_k}|^2$.  This norm of course
induces a metric, namely, a way of defining the distance between
tensors of the same rank.  It is easy to see that the norm of a
tensor $\BC$ is equal to the Eucledian norm of its corresponding
state $\ket{\BC}$. Also, for a rank-2 tensor (which can be viewed
as a matrix), it is known that its \emph{operator} norm is not
larger than its tensor norm (which in this case is simply the
Frobenious norm).  

It is true that for any three tensors, $\BB_1,\BB_2, \BB$, we have
$\|\BB_1\BB-\BB_2 \BB\|\le \|\BB_1-\BB_2\|\cdot\|\BB\|$.  In fact,
many times in the context of MPSs, a much stronger inequality 
holds. Assume $\BB$ connects with $\BB_1$ or $\BB_2$ along one edge,
indexed by $\alpha$. Assume further that $\|\BB_\alpha\|=1$ for
every $\alpha$ (in the context of canonical MPSs, it will often be
the case that we consider the contraction of one side of the chain
with a fixed index $\alpha$ of the cut edge, and this contraction is
indeed of norm $1$ by the canonical conditions).  In this case, we
have a much stronger inequality, which can be easily verified: 
\begin{align}
\label{eq:contract} 
  \norm{\BB_1\BB-\BB_2 \BB}= \norm{\BB_1-\BB_2} \ .
\end{align}
 
We can apply this to cases of interest, when we compare contractions
of tensors which differ in only a single term. For example, consider
vector $\lambda_{\alpha}$ with norm $1$ and two tensors
$B_{\alpha,i_1,\ldots,i_k}$ and $A_{\alpha,j_1,\ldots,j_l}$ such
that when $\alpha$ is fixed, the resulting tensors $\bm{A}_{\alpha}$
and $\bm{B}_{\alpha}$ have norm 1. Let $\hat{\Bl}$, $\hat{\bm{A}}$
and $\hat{\bm{B}}$ be tensors with the same rank and dimensions as
$\Bl$, $\bm{A}$ and $\bm{B}$. We have, by \Eq{eq:contract}, 
\begin{align}
\label{eq:dist1}
  \norm{\bm{A}\Bl\bm{B}  - \bm{A}\hat{\Bl}\bm{B}} 
    = \norm{\Bl - \hat{\Bl}} \ ,
\end{align}
and also 
\begin{align}
\label{eq:dist1a}
  \norm{\bm{A}\Bl\bm{B}  - \hat{\bm{A}}\hat{\Bl}\bm{B}} 
    = \norm{\bm{A} \Bl - \hat{\bm{A}} \hat{\Bl}} \ .
\end{align}
And similarly,
\begin{align}
  \nonumber
  \norm{\bm{A}\Bl\bm{B}  - \bm{A}\Bl\hat{\bm{B}}} 
  &= \norm{\Bl\bm{B}  - \Bl\hat{\bm{B}} } \\
  &= \left[ \sum_{\alpha} |\lambda_{\alpha}|^2
     \norm{( \bm{B}_{\alpha} - \hat{\bm{B}}_{\alpha} )}^2 
        \right]^{1/2} \nonumber\\
  &\le \max_{\alpha} \| \BB_{\alpha} - \hat{\BB}_{\alpha} \| \ .
\label{eq:dist2}
\end{align}

%%%%%%%%%%%%%%%%%%%%%%%%%%%%%%%%%%%%%%%%%%%%%%%%%%%%%%%%%%%%%%%%%%%%%%%%%%5
\section{The Algorithm}
\label{sec:Algorithm}

As discussed earlier, in order to carry out the outline described in
\Sec{sec:results}, we would like to work with canonical MPSs.
Additionally, since the tensor pairs $(\Bl, \BB)$ for neighboring
nodes overlap, we would like an $\epsilon$ net over canonical pairs
such that $\Bm(\Bl,\BB)$ could be equal to the $\Bl$ of the next
pair (we call this \emph{ perfect stitching}). We do not know how to
efficiently construct an $\epsilon$ net that satisfies those
conditions exactly; we resort to approximately canonical MPSs with
approximate stitching. 

%
% --------------------------------------------------------------------
\subsection{ $\epsilon$ nets}

We fix $\epsilon>0$ (to be determined later) and define two
$\epsilon$ nets.  We start with discretizing $\BG^{[1]}$ and
$\BG^{[n]}$.
\begin{definition}[the $G_{end}$ $\epsilon$ net]
\label{def:G-end} 
  $G_{end}$ is a set of canonical boundary tensors [see
  \Eq{eq:cond-bound}] such that, for any canonical boundary
  tensor $\hat{\BG}$ there is $\BG \in G_{end}$ such that for each
  $\alpha$, $\|\hat{\BG}_{\alpha} - \BG_{\alpha}\| \le \epsilon$. 
\end{definition}
We now define an $\epsilon$ net over the intermediate tensors, 
or more precisely, for the pairs $(\Bl, \BB)$.  
\begin{definition}[the $G$ $\epsilon$ net]
\label{def:G} 
  $G$ is a set of pairs of tensors $(\Bl,\BB)$
  such that:
  \begin{enumerate}
  
    \item \textbf{$\Bl$ is positive and normalized:} For all $\alpha$ 
    $\lambda_\alpha\ge 0$ and $\braket{\Bl}{\Bl} =1.$
      
    \item \textbf{$G$ is an $\epsilon$ net:} 
       For every canonical triplet $(\hat{\Bl}, \hat{\BG},
       \hat{\Bm})$ there is $(\Bl, \BB)\in G$ such 
        $\|\hat{\Bl}\hat{\BG}\hat{\Bm}-\Bl\BB\|\le \epsilon$. 
  
    \item \textbf{$\BB$ is \textbf{perfectly right canonical}:}
      For every $\alpha,\alpha'$, $\langle{\BB_{\alpha}}|\BB_{\alpha'}\rangle=
      \delta_{\alpha\alpha'}$ (here $\alpha,\alpha'$ are the 
      left Greek indices of $\BB$).
      
    \item \textbf{$(\Bl,\BB)$ are \textbf{approximately left canonical}:}    
      For every $\beta\neq \beta'$,
      \begin{align}
        \label{eq:Lcan}
        |\langle{(\Bl\BB)_{\beta}}|(\Bl\BB)_{\beta'}\rangle|
                  \le 3 \epsilon \ .
      \end{align}

  \end{enumerate}
\end{definition}

% --------------------------------------------------------------------
\subsection{$\epsilon$ net Generators}
\label{sec:generators}

We now explain how to construct such nets efficiently.  Both
generators for the $\epsilon$ nets will make use of the following
general lemma
\begin{lemma}
\label{lem:general-generator} 
  For any positive integers $a \le b$ and any $\nu$ in the range
  $\left(0,1/\sqrt{a} \right]$, we can generate a set of $a \times
  b$ matrices $S_{ab}$ over the complex numbers such that for any
  $A \in S_{ab}$, the rows of $A$ are an ortho-normal set of length
  $b$ vectors. Furthermore, for any $a \times b$ matrix $B$ whose
  rows form a set of orthonormal vectors, there is a matrix $A \in
  S_{ab}$ such that each row of $A-B$ has $L_2$ norm at most $\nu$.
  The size of $S_{ab}$ is at most $\bigO((72 b/\nu)^{2ab})$.  The
  time to generate $S_{ab}$ is $\bigO(a^2 b (72 b/\nu)^{2ab} )$. If
  $a=1$, we can generate a set of vectors with real non-negative
  entries, rather than complex.  The size of the net is $\bigO((72
  b/\nu)^{b})$ and the time to generate it is $\bigO( b (72
  b/\nu)^{b})$.
\end{lemma}

The proof appears in the Appendix.

\vspace{.1in}

\subsubsection{Generating $G_{end}$:}
\label{sec:endnet}

%By \Eq{eq:end-cond}, every $\Gamma \in G_{end}$ is essentially an
%isometry $\mathcal{C}^d \mapsto \mathcal{C}^D$. We can therefore
%generate it by looking at a $D\times D$ unitary matrix
%$U_{\alpha\beta}$, and identifying $\Gamma^i_\alpha \EqDef
%U_{i\alpha}$. We can therefore generate an $\epsilon$ net over the
%$D\times D$ unitaries and this will give us an $\epsilon$ net over
%the $\Gamma$'s. 

% One can easily construct such a net which contains 
%$N_{end}=\mathcal{O}\Big((poly(D)/\epsilon)^{D^2}\Big)$ elements.
%For example, a crude way to do this is  
%by generating a finer net on the $D\times D$ 
%matrices with entries bounded by $1$ in absolute value, by discritizing 
%each entry independently to intervals of length $\epsilon/poly(D)$. 
%Then for every matrix, compute
%the $\ell_2$ distance from unitarity, discard those matrices 
%that are too far, 
%and apply Grahm-Schmidt on the remaining matrices to make them 
%perfectly unitary.  
%The construction of such a net takes time $\mathcal{O}(N_{end})$
%times $poly(D)$.  

%Recall, that we assume that $d \le D$. 

Invoke Lemma~\ref{lem:general-generator} with $\nu=\epsilon$, $a=D$,
and $b=d_{end}$. For every $A \in S_{D,d_{end}}$, add a $\BG$ to the
$\epsilon$ net, where $A_{\alpha,i} = \Gamma^i_{\alpha}$. Note that
the conditions of Lemma \ref{lem:general-generator}, are satisfied
if $\epsilon \le 1/\sqrt{D}$. Since $d_{end} \le Dd$, the size of
the net is at most $(72 Dd/ \epsilon)^{2dD^2}$ and the time to
generate it is $O(d D^3)$ times the size of the set.

%
%++++++++++++++++++++++++++++++
%
\subsubsection{Generating $G$:}

We generate $G$ by first generating an $\epsilon/2$-net over the
$\Bl$'s and an $\epsilon/2$-net over the $\BB$'s. To generate the
net of the $\Bl$'s, invoke Lemma \ref{lem:general-generator} with
$a=1$, $b=D$ and the $\nu$ in the lemma set to $\epsilon/2$. Note
that we would like to have a $\Bl$ with non negative real entries.
According to Lemma~\ref{lem:general-generator}, this actually
requires fewer items in our net since we are omitting the phases in
each entry in the tensor. The resulting net for the $\Bl$'s has size
$(144 D/\epsilon)^D$ and can be generated in time $O(D(144
D/\epsilon)^D)$. 

To generate the net over the $\BB$'s, we invoke Lemma
\ref{lem:general-generator} with $a=D$, $b=dD$, and $\nu=\epsilon/2$.
Note that in order to invoke Lemma~\ref{lem:general-generator}, we
require that $\epsilon \le 2/\sqrt{D}$. For any matrix
$A_{\alpha,(i,\beta)}$ in the set, we generate a tensor $\BB$ where
$B^i_{\alpha,\beta} = A_{\alpha,(i,\beta)}$. This way we generate a
set of pairs $(\Bl,\BB)$ which satisfies both the normalization
condition [condition~(1) of \Def{def:G}] and the condition of being
perfect right canonical [condition~(3) of \Def{def:G}].

To see that we in fact have an $\epsilon$ net [i.e. condition~(2)
is satisfied], consider a perfectly canonical pair $(\Bl, \BB)$,
and let us find a pair $(\hat{\Bl},\hat{\BB})$
in the net that is $\epsilon$-close to it. We
first replace $\Bl$ with a $\hat{\Bl}$ from the first net and then
replace $\BB$ with a $\hat{\BB}$ from the second net. 
Using \Eq{eq:dist1}, we have that
\begin{align*}
  \norm{\Bl \BB - \hat{\Bl}\BB}
    = 
  \norm{\Bl  - \hat{\Bl}}
    \le  \frac{\epsilon}{2}  \ ,  
\end{align*}
Using \Eq{eq:dist2}, we also have 
\begin{align*}
  \norm{\hat{\Bl} \BB - \hat{\Bl}\hat{\BB} } 
    \le \max_\alpha \norm{ \BB_{\alpha} - \hat{\BB}_{\alpha} }  
    \le \frac{\epsilon}{2} \ .  
\end{align*}

Next, we discard all tensors $(\Bl,\BB)$ that are not
approximately left canonical, namely, those that violate condition $(4)$. 
It remains to show that the remaining tensors still satisfy
condition $(2)$, that is, the $\epsilon$ net condition. We do that by
showing that a pair $(\Bl, \BB)$ that is $\epsilon$ close to a
canonical triplet must necessarily be approximately left canonical.
Therefore, such a pair would not have been eliminated.

To see this, let the tensor $\BA=\Bl\BG\Bm$ be the contraction of
the canonical triplet and $\BC$ be the contraction of $\hat{\Bl}
\hat{\BB}$ from the net such that $\norm{\BA-\BC}\le \epsilon$.  The
fact that $\BA$ is perfectly left canonical is expressed in the fact
that for every $\beta\neq\beta'$,
$\bra{\BA_\beta}\BA_{\beta'}\rangle =0$. To prove that $\BC$ is
approximately left canonical, we need to show
$|\bra{\BC_\beta}\BC_{\beta'}\rangle| \le 3\epsilon$. Indeed, 
$\norm{\BA-\BC}\le \epsilon$ implies $\norm{\BA_{\beta} -
\BC_{\beta}} \le \epsilon$ for every $\beta$.  Assume $\beta\neq
\beta'$. Then
\begin{align*}
  |\braket{ \BC_{\beta} }{ \BC_{\beta'} }|
    &=  |\braket{\BA_{\beta} + (\BC_{\beta} - \BA_{\beta})}
            { \BA_{\beta'} + (\BC_{\beta'} - \BA_{\beta'})}| \\
   &\le |\braket{ \BA_{\beta} }{\BA_{\beta'}} | 
     + |\braket{\BA_{\beta}}{\BC_{\beta'} - \BA_{\beta'} }| \\
     &\ \ + \ |\braket{ \BC_{\beta} - \BA_{\beta} }{\BA_{\beta'}} |
     + |\braket{\BC_{\beta} - \BA_{\beta}  }{\BC_{\beta'} - \BA_{\beta'}  }| \\
   &\le  \norm{\BA_{\beta}} \norm{\BC_{\beta'} - \BA_{\beta'} } 
     + \norm{\BA_{\beta'}} \norm{\BC_{\beta} - \BA_{\beta} } \\
     &\ + \ \norm{\BC_{\beta} - \BA_{\beta} } \norm{\BC_{\beta} - \BA_{\beta} } \\
   &\le  2 \epsilon + \epsilon^2 \le 3 \epsilon \ .
\end{align*}
This concludes the proof that $G$ is indeed an $\epsilon$ net
according to \Def{def:G}.

%
%+++++++++++++++++++++++++++++++++++++++++++++++++++++++++
%
\subsubsection{Complexity of Generating $G$ and $G_{end}$:}

By \Lem{lem:general-generator}, $N \EqDef |G|$, the size of the
$\epsilon$ net $G$ is
\begin{align}
\label{eq:N}
  N = \mathcal{O} \left(\frac{144dD}{\epsilon}\right)^{D + 2dD^2}. 
\end{align}
This is the size of the set formed by taking all pairs $(\Bl,\BB)$,
where each $\Bl$ and $\BB$ come from their respective nets. The time
required to generate the original net (before tensors are discarded)
is $O(dD^3N)$. The cost of checking whether a $(\lambda,B)$ pair is
approximately left canonical is $\bigO(dD^3)$, so the total cost of
generating the net is $\bigO(dD^3N)$.

For $G_{end}$, both the number of points and the running time 
which were determined in \Sec{sec:endnet},
are bounded above by the corresponding bounds of $G$.

% --------------------------------------------------------------------
\subsection{The algorithm} 

When processing particle $j$, the algorithm creates a list $L_j$
of partial solutions, one for each $(\Bl,\BB)$ pair in $G$.
For each such partial solution, a tail (i.e., the tensors to the
left of the $j$th particle)
and energy is kept.

\begin{description}
  \item[First step:]  
    Create the first list $L_2$: For each  $(\Bl^{[2]},
    \BB^{[2]})\in G$, find its tail, namely the $\BG^{[1]}\in
    G_{end}$ which minimizes the energy with respect to $H_{1,2}$ of
    the tensor $\BG^{[1]} \Bl^{[2]} \BB^{[2]}$. Denote this minimal energy by 
    $E_2(\Bl^{[2]}, \BB^{[2]})$. We keep
    both the tail and the computed energy, for each pair $(\Bl^{[2]},
    \BB^{[2]})\in G$. 
    
  \item [Going from $j=3$ to $j=n-1$:]
    we assume we have created the list $L_{j-1}$.  For each pair
    $(\Bl^{[j-1]}, \BB^{[j-1]}) \in G$ there is a tail in $L_{j-1}$:
    \begin{align*}
      \BG^{[1]}, (\Bl^{[2]},\BB^{[2]}),
        (\Bl^{[3]},\BB^{[3]}),\ldots,(\Bl^{[j-2]},\BB^{[j-2]})
    \end{align*}
    and an energy value that we denote by $E_{j-1}(\Bl^{[j-1]},
    \BB^{[j-1]})$. To create $L_j$, we find a tail for each
    $(\Bl^{[j]}, \BB^{[j]}) \in G$. We
require that the tail for a given $(\Bl^{[j]}, \BB^{[j]})$ is an
    item in $L_{j-1}$ which satisfies the ``\textbf{\em
    stitching}'' condition:
    \begin{align}
      \label{eq:stitching}
      \norm{ \Bm(\Bl^{[j-1]}, \BB^{[j-1]}) - \Bl^{[j]}} \le
      2\epsilon \ .
    \end{align}
    
    We pick the tail for $(\Bl^{[j]}, \BB^{[j]})$ to be an item in $L_{j-1}$
which satisfies the stitching condition and minimizes 
    $H_{j-1,j}(\Bl^{[j-1]} \BB^{[j-1]} \BB^{[j]})+
    E_{j-1}(\Bl^{[j-1]}, \BB^{[j-1]})$. The minimum such value is defined to be
    $E_j(\Bl^{[j]}, \BB^{[j]})$. 

  \item [Final step:] 
    The final step, $j=n$, is exactly as in the intermediate steps
    except the algorithm goes over $ \BG^{[n]} \in G_{end}$, rather
    than over pairs from $G$ and there is no stitching constraint. 
    More precisely, we pick the tail for $ \BG^{[n]}$ to be the item
    in $L_{n-1}$ which minimizes $H_{n-1,n}( \Bl^{[n-1]} \BB^{[n-1]}
    \BG^{[n]}) + E_{n-1}(\Bl^{[n-1]}, \BB^{[n-1]})$. The minimal
    value is defined to be $E_n( \BG^{[n]} )$.

    Finally, we choose $\BG^{[n]}$ which minimizes $E_n(\BG^{[n]})$. We
    output the MPS that is defined by $\BG^{[n]}$ and its tail:
    \begin{align}      
      \ket{\Omega} \EqDef \ket{\BG^{[1]} \Bl^{[2]} \BB^{[2]}
       \BB^{[3]}  \cdots \BB^{[n-1]}\BG^{[n]}} \ ,
    \end{align}
    together with the energy which the algorithm calculated: 
    \begin{align}
      E_{alg} (\Omega)  \EqDef E_n(\BG^{[n]}) \ . 
    \end{align}
\end{description}

Note that since each $(\Bl^{[j]},\BB^{[j]})$ is perfectly
right canonical, the state $\ket{\Omega}$ is normalized.  This can be seen
by contracting the tensor network corresponding to the inner product
$\braket{\Omega}{\Omega}$ from right to left.  

Unlike in the classical case, our algorithm does not search all
states due to the discretization. Moreover, it does not optimize
over the real energy of the states that it does check, but rather
over $E_{alg}(\Omega) =\sum_{j}
H_{j-1,j}(\Bl^{[j-1]}\BB^{[j-1]}\BB^{[j]})$. $E_{alg}$ is different
from the true energy $E$ because the states are not exactly
canonical.  Note that the output $E_{alg}(\Omega)$ is thus just an
approximation of the real energy $E(\Omega)$ of the output MPS
$\ket{\Omega}$. We output $E_{alg}(\Omega)$ anyway, since our
guarantee on its error is somewhat better than on the error for
$E(\Omega)$, as we will see in \Sec{sec:Error}.

The following claim easily follows from the same
reasoning as for the classical dynamic programming algorithm: 
  
\begin{claim}
  \label{cl:min} The algorithm finds the state which minimizes
    $E_{alg}$ among all MPSs of the form $\BG^{[1]} \Bl^{[2]}
    \BB^{[2]} \BB^{[3]} \cdots \BB^{[n-1]} \BG^{[n]}$, such that
    $\BG^{[1]},\BG^{[n]}\in G_{end}$,  
    $(\Bl^{[j]}, \BB^{[j]})\in G $ for all 
$j\in\{2,...,n-1\}$, and the stitching conditions
    (\Eq{eq:stitching}) are all satisfied.
\end{claim}

%
% ======================================================================
%

\section{Error and Complexity Analysis}
\label{sec:Error}

In order to finish the proof of \Thm{thm:mps}, we will prove
the theorem below. As noted above, this theorem actually gives a
better error bound on $E_{alg}(\Omega)$ than the bound on
$E(\Omega)$ that is given in \Thm{thm:mps}.
\begin{theorem}[Error bound]
\label{thm:main} 
  Let $E_0$ be the minimal energy that can be achieved
  by a state with bond dimension $D$, and $J$ the maximal operator norm $\|H_{j,j+1}\|$
over all terms. Then:  
  \begin{align}
  \label{eq:main}
    E_{alg}(\Omega) - 6Jn\epsilon \le E_0 \le 
      E(\Omega) \le E_{alg}(\Omega) + \tfrac{3}{2}JD^2n^2\epsilon  \ .
  \end{align}
\end{theorem} 
It is easy to verify that as long as $nD^2\ge 12$, \Eq{eq:main} implies
\Eq{eq:mps} of \Thm{thm:mps}.

\begin{proof}

  By definition, $E_0 \le E(\Omega)$.  We first prove that
  $E_{alg}(\Omega) - 6Jn\epsilon \le E_0$. Let:
  \begin{align*}
    \ket{\psi} = \ket{\hat{\BG}^{[1]}
      \hat{\Bl}^{[2]}\hat{\BG}^{[2]} \cdots
      \hat{\Bl}^{[n]}\hat{\BG}^{[n]}} \ .
  \end{align*}
  be a state with $E(\psi) = E_0$ of bond dimension $D$, written as a canonical
  MPS.  For every triplet $( \hat{\Bl}^{[j] } ,\hat{\BG}^{[j]},
  \hat{\Bl}^{[j+1]})$ for $j=2,\ldots n-1$, we associate a pair
  $(\tilde{\Bl}^{[j]}, \tilde{\BB}^{[j]}) \in G$ which is
  $\epsilon$-close to that triplet.  In addition, we find
  $\tilde{\BG}_1\in G_{end}$ close to $\hat{\BG}_1$ and
  $\tilde{\BG}_n\in G_{end}$ close to $\hat{\BG}_n$. We define the
  state:
  \begin{align*}
    \ket{\phi} = \ket{\tilde{\BG}_1 \tilde{\Bl}_2 \tilde{\BB}_2
      \tilde{\BB}_3 \cdots \tilde{\BB}_{n-1}\tilde{\BG}_n} \ .
  \end{align*}

  Just like $\ket{\Omega}$, this state is normalized due to the 
  fact that the tensors in $G_{end}$ and $G$ are perfectly
  right canonical.

  We first claim that $E_{alg}(\Omega) \le E_{alg}(\phi)$. This
  follows from the fact that $\ket{\phi}$ belongs to the set of
  states over which the dynamic algorithm searches (see
  Claim~\ref{cl:min}), since the
  $\tilde{\Bl}^{[j-1]}\tilde{\BB}^{[j-1]}$ and 
  $\tilde{\Bl}^{[j]}\tilde{\BB}^{[j]}$ satisfy the stitching
  condition~(\ref{eq:stitching}), as promised by the following
  lemma:
  \begin{lemma}
    For every $j=3, \ldots, n-1$,
    \begin{align}
      \norm{\mu(\tilde{\Bl}^{[j-1]},
        \tilde{\BB}^{[j-1]}) - \tilde{\Bl}^{[j]}}\le 2\epsilon \ . 
    \end{align}
  \end{lemma}
  \begin{proof}
    We use the fact (established in Lemma \ref{lem:Schmidt} in the
    Appendix) that for any two bipartite states $\ket{A} = \sum_i
    a_i\ket{i}\ket{A_i}$, with normalized $\ket{A_i}$, $\ket{B} =
    \sum_i b_i\ket{i}\ket{B_i}$ with normalized $\ket{B_i}$, we have
    $\sum_i |a_i-b_i|^2\le \|A-B\|^2.$

    The tensors $\hat{\Bl}^{[j]}\hat{\BG}^{[j]}\hat{\Bl}^{[j+1]}$
    and $\tilde{\Bl}^{[j]} \tilde{\BB}^{[j]}$ represent two quantum
    states on 3 particles, where in both states, the Schmidt basis
    of the \emph{first} particle is the standard basis, and the
    perfect right canonical condition of Definition \ref{def:G} (or
    alternatively, the condition of Equation \ref{eq:cond-right})
    holds.  The Schmidt coefficients are given by
    $\{\hat{\lambda}^{[j]}_\alpha\}$ and
    $\{\tilde{\lambda}^{[j]}_\alpha\}$, respectively.  According to
    the above fact (Lemma \ref{lem:Schmidt}) 
    \begin{align}
    \label{eq:small-lam}
      \norm{\hat{\Bl}^{[j]} - \tilde{\Bl}^{[j]}} 
      \le \norm{\hat{\Bl}^{[j]}\hat{\BG}^{[j]}
        \hat{\Bl}^{[j+1]}
        - \tilde{\Bl}^{[j]} \tilde{\BB}^{[j]}} 
      \le \epsilon \ .
    \end{align}

    Similarly, we know that
    $\norm{\hat{\Bl}^{[j-1]}\hat{\BG}^{[j-1]}\hat{\Bl}^{[j]}
    - \tilde{\Bl}^{[j-1]} \tilde{\BB}^{[j-1]}} \le \epsilon$.
    Consider now these 3-particle states expanded in terms of the
    basis vectors $\ket{\beta}$ of the \emph{third} particle.
    Denote these expansions by $\sum_{\beta}
    a_{\beta}\ket{v_{\beta}}\ket{\beta}$, with normalized
    $\ket{v_{\beta}}$, and $\sum_{\beta}
    b_{\beta}\ket{w_{\beta}}\ket{\beta}$ with normalized
    $\ket{w_{\beta}}$, respectively. Then by definition, 
    $a_{\beta}=\hat{\lambda}^{[j]}_\beta$, and 
    $b_{\beta}=\mu_\beta(\tilde{\Bl}^{[j-1]},
    \tilde{\BB}^{[j-1]})$. We can therefore apply again Lemma
    \ref{lem:Schmidt} and get: $\norm{\mu(\tilde{\Bl}^{[j-1]},
    \tilde{\BB}^{[j-1]}) - \hat{\Bl}^{[j]}}\le \epsilon$.
    Together with \Eq{eq:small-lam}, we therefore
    obtain $\norm{\mu(\tilde{\Bl}^{[j-1]}, \tilde{\BB}^{[j-1]}) -
    \tilde{\Bl}^{[j]}}\le 2\epsilon$. 
  \end{proof}

  Thus far, we have established that $E_{alg}(\Omega) \le
  E_{alg}(\phi)$. We will therefore prove the inequality
  $E_{alg}(\Omega) - 6Jn\epsilon \le E_0$ by showing that
  $|E_{alg}(\phi) - E_0|\le 6nJ\epsilon$.  Observe that each energy
  term in $E_0$ depends solely on two overlapping triplets
  $\hat{\Bl}^{[j]}\hat{\BG}^{[j]}\hat{\Bl}^{[j+1]}
  \hat{\BG}^{[j+1]}\hat{\Bl}^{[j+2]}$. The corresponding energy term
  in $E_{alg}(\phi)$ depends only on 
  $\tilde{\Bl}^{[j]}\tilde{\BB}^{[j]}\tilde{\BB}^{[j+1]}$. We now
  bound the distance between these two tensors.  We have 
  \begin{align*}
    & \tilde{\Bl}^{[j]}\tilde{\BB}^{[j]}\tilde{\BB}^{[j+1]}
     - \hat{\Bl}^{[j]}\hat{\BG}^{[j]} \hat{\Bl}^{[j+1]}
     \hat{\BG}^{[j+1]} \hat{\Bl}^{[j+2]} \\
    &=\big(\tilde{\Bl}^{[j]}\tilde{\BB}^{[j]} - 
    \hat{\Bl}^{[j]}\hat{\BG}^{[j]}\hat{\Bl}^{[j+1]}) 
      \tilde{\BB}^{[j+1]} \\
    &+\hat{\Bl}^{[j]}\hat{\BG}^{[j]}
      \big(\hat{\Bl}^{[j+1]} - \tilde{\Bl}^{[j+1]}\big)
        \tilde{\BB}^{[j+1]} \\
    &+  \hat{\Bl}^{[j]}\hat{\BG}^{[j]}
      \big(\tilde{\Bl}^{[j+1]}\tilde{\BB}^{[j+1]} -
      \hat{\Bl}^{[j+1]}
     \hat{\BG}^{[j+1]}\hat{\Bl}^{[j+2]}\big) \
  \end{align*}    

  Taking the LHS and RHS sides of the above equation, and using
  \Eq{eq:dist1} and \Eq{eq:dist1a}, we have that
  \begin{align*}
    & \norm{\tilde{\Bl}^{[j]}\tilde{\BB}^{[j]}\tilde{\BB}^{[j+1]}
     - \hat{\Bl}^{[j]}\hat{\BG}^{[j]} \hat{\Bl}^{[j+1]}
     \hat{\BG}^{[j+1]} \hat{\Bl}^{[j+2]}} \\
    &\le \| \tilde{\Bl}^{[j]}\tilde{\BB}^{[j]} - 
    \hat{\Bl}^{[j]}\hat{\BG}^{[j]}\hat{\Bl}^{[j+1]} \| \\
    &+ \norm{\hat{\Bl}^{[j+1]} - \tilde{\Bl}^{[j+1]}} \\
    &+ \norm{\tilde{\Bl}^{[j+1]}\tilde{\BB}^{[j+1]} -
      \hat{\Bl}^{[j+1]}
     \hat{\BG}^{[j+1]}\hat{\Bl}^{[j+2]}} \ .
  \end{align*}    
  The first and third term in the above sum can be bounded by
  $\epsilon$ because of the condition of the $\epsilon$ net $G$. The
  norm of the middle term is bounded in \Eq{eq:small-lam}. Therefore
  the norm of the difference between the tensors is at most
  $3\epsilon$.  It follows that the difference between the two
  energy contributions is at most $6\epsilon\norm{H_{j,j+1}} \le
  6\epsilon J$. 

  We illustrate the boundary cases by working through the analysis
  for the left end of the chain. We want to bound $\|
  \hat{\BG}^{[1]} \hat{\Bl}^{[2]} \hat{\BG}^{[2]} \hat{\Bl}^{[3]} - 
  \tilde{\BG}^{[1]} \tilde{\Bl}^{[2]}\tilde{\BB}^{[2]} \|.$ Note
  that $\| \hat{\BG}^{[1]} (\hat{\Bl}^{[2]} \hat{\BG}^{[2]}
  \hat{\Bl}^{[3]}-\tilde{\Bl}^{[2]}\tilde{\BB}^{[2]} ) \| $ is
  bounded by $\epsilon$ because of the conditions on the
  $\epsilon$ net and \Eq{eq:contract}. Using \Eq{eq:dist2}, we have
  that
  \begin{align*}
     \norm{(\hat{\BG}^{[1]} - \tilde{\BG}^{[1]}) 
       \tilde{\Bl}^{[2]} \tilde{\BB}^{[2]}}
     \le \max_{\alpha} 
       \norm{ \hat{\BG}^{[1]}_{\alpha} - \tilde{\BG}^{[1]}_{\alpha}}
     \le \epsilon \ . 
  \end{align*}
  Hence, the overall bound on the difference is $2\epsilon$.  It
  follows that the difference between the two energy contributions
  is it most $4 \epsilon\norm{H_{1,2}} \le 4\epsilon J$. A similar
  argument holds for $H_{n-1,n}$. \vspace{.1in}

  Now we turn to the right inequality in \Thm{thm:main} and
  show $|E(\Omega)-E_{alg}(\Omega)|\le \tfrac{3}{2}JD^2n^2\epsilon$.
  We bound the difference in energy for each term $H_{j-1,j}$. The
  contribution of this term to $E_{alg}(\Omega)$ is calculated from 
  $\Bl^{[j-1]}\BB^{[j-1]}\BB^{[j]}$. The true energy, however,
  depends on $\BG^{[1]}\Bl^{[2]}\BB^{[2]}\BB^{[3]}\cdots \BB^{[j]}$
  since $\ket{\Omega}$ is only approximately left canonical. We will
  show that the error accumulates linearly as we sweep from left to
  right, summing up to $3jJD^2\epsilon$ for $H_{j-1,j}$. Therefore,
  the total error is $| E_{alg}(\Omega) - E(\Omega) | \le
  \tfrac{3}{2}JD^2n^2\epsilon$.

  We now provide a more accurate argument.  The energy estimate for
  the term $H_{j-1,j}$ is calculated from the contraction
  $\Bl^{[j-1]}\BB^{[j-1]}\BB^{[j]}$. Graphically, this contribution
  is given by
  \begin{center}
    \includegraphics[scale=1]{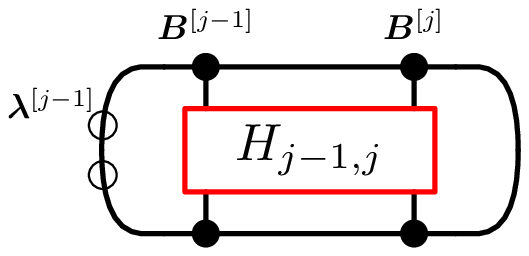}
  \end{center}
  The true energy, however, is calculated from the contraction of
  $\Bl^{[2]}\BB^{[2]}\BB^{[3]}\cdots \BB^{[j]}$. Graphically, this
  is given by
  \begin{center}
    \includegraphics[scale=0.75]{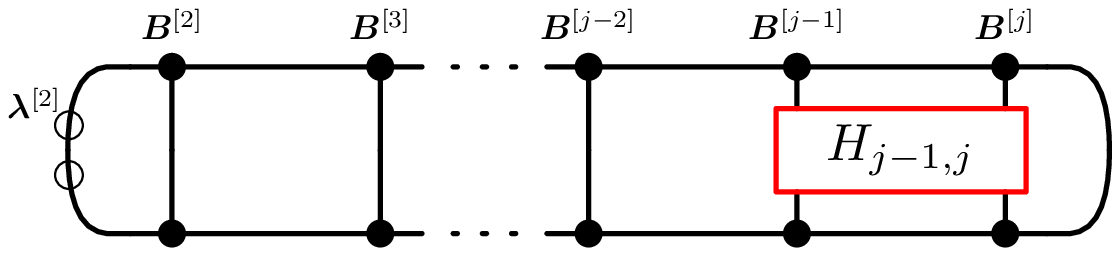}
  \end{center}
  (Notice that we have collapsed the $\BG^{[1]}$ terms because of
  the canonical condition~\ref{eq:cond-bound} -- see
  \Fig{fig:cond} (b)).

  Had the state $\ket{\Omega}$ been perfectly left canonical, the
  two would have been the same. But since it is only approximately
  canonical from the left, there is some difference that can be
  bounded. The analysis is done iteratively from left to right. We
  start by writing
  \begin{center}
    \includegraphics[scale=0.7]{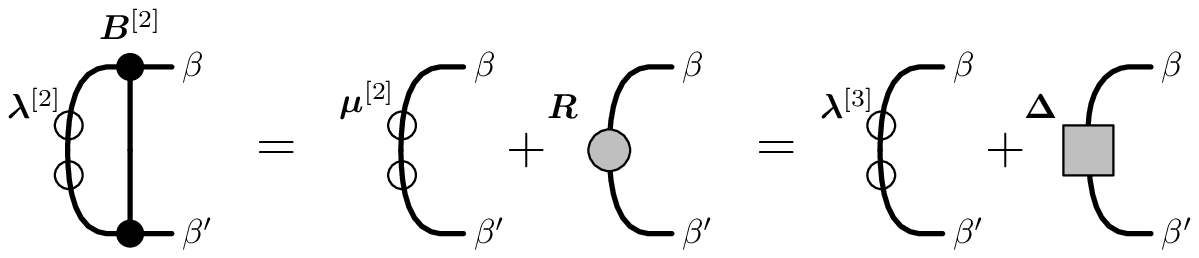} 
  \end{center}
  In this picture, the tensor $R_{\beta\beta'}$ is defined to be off diagonal
  (i.e., equal to zero on the diagonal: $R_{\beta\beta}=0$) and for
  the $\beta\neq\beta'$ terms, it is defined by
  $R_{\beta\beta'}=\braket{(\Bl^{[2]}\BB^{[2]})_{\beta}}{(\Bl^{[2]}\BB^{[2]})_{\beta'}}=
  \sum_{\alpha, i} |\lambda^{[2]}_\alpha|^2 B^{[2]i}_{\alpha\beta}
  (B^{[2]i}_{\alpha\beta'})^*$.  $\BDel$ is defined by:
  \begin{align*}
    \Delta_{\beta\beta'} \EqDef
      R_{\beta\beta'} + \delta_{\beta\beta'}(|\lambda^{[3]}_\beta|^2 -
        |\mu^{[2]}_\beta|^2) \ .
  \end{align*}
  Using the fact that $(\Bl^{[2]}, \BB^{[2]})$ is approximately
  left canonical (see \Eq{eq:Lcan}), and the stitching conditions of
  $\Bl^{[3]}$ and $\Bm^{[2]}$ (see \Eq{eq:stitching}), it is easy to
  see that for every $\beta,\beta'$, 
  \begin{align}
    \label{eq:Delta-bound}  
    |\Delta_{\beta\beta'}| \le 3\epsilon \ .
  \end{align}
  We may therefore write the true energy contribution as the sum of 
  \begin{center}
    \includegraphics[scale=0.7]{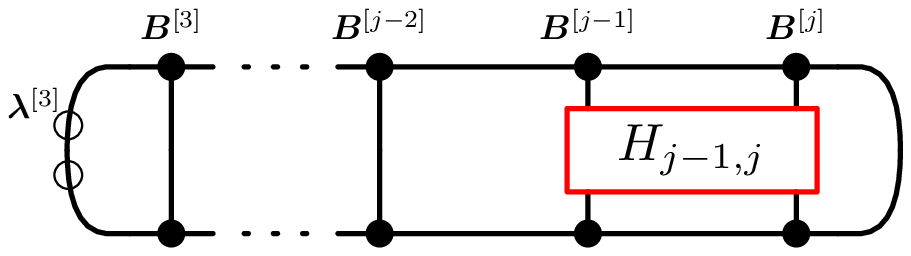} 
  \end{center}
  and 
  \begin{center}
    \includegraphics[scale=0.7]{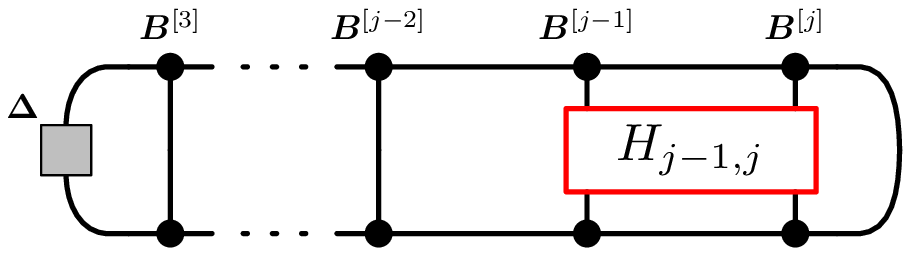} 
  \end{center}
  The analysis of the first term is done in the next iteration step. 
  The second term can be seen as the error introduced by the fact
  that $(\Bl^{[2]}\BB^{[2]})$ is approximately left canonical. To
  estimate its size, notice that it can be viewed as the expectation
  value of the operator $\BDel\otimes H_{j-1,j}$ (here $\BDel$ is
  viewed as a matrix), with respect to the MPS that is described by
  $\ket{\BB^{[3]}\BB^{[4]}\cdots \BB^{[j]}}$.  Using \Eq{eq:Delta-bound}
  and the assumption $\norm{H_{j-1,j}}\le J$, it is easy to see that
  $\norm{\BDel \otimes H_{j-1,j}} \le 3JD\epsilon$. Here, in both
  cases, we used $\norm{\cdot}$ to denote the \emph{operator} norm
  of $\bm{\Delta} \otimes H_{j-1,j}$, instead of the
  usual tensor norm; we can do this since the operator norm is at
  most as large as the tensor norm, and the tensor norm of
  $\bm{\Delta}$ is at most $3D\epsilon$.  Moreover, the norm of the
  MPS $\ket{\BB^{[3]}\BB^{[4]}\cdots \BB^{[j]}}$ is exactly $\sqrt{D}$
  (it would have been exactly 1 had there been a $\Bl^{[3]}$ term
  before $B^{[3]}$), and therefore the amplitude of second term is
  upper bounded by $3JD^2\epsilon$. 
  
  Carrying the same analysis all way to $(\Bl^{[j-2]},
  \BB^{[j-2]})$, we end up with a term that is identical to the
  energy estimation of the algorithm, plus some error term whose
  amplitude is at most $3jJD^2\epsilon$. Therefore, by  simple
  algebra, we have that for
  the total system,
  \begin{align}
    | E_{alg} - E(\Omega) | \le \tfrac{3}{2}JD^2n^2\epsilon \ .
  \end{align}

\end{proof}

For a target error $\delta$, we select $ \epsilon \le
\frac{\delta}{2JD^2n^2} \ .$ Using the bound from \Eq{eq:N}, we get
that the size of the net for the interior particles is
\begin{align}
  N = \bigO \left( 
    \left(\frac{144 JdD^3n^2}{\delta}\right)^{D + 2dD^2} \right).
\end{align}

Note that in using Lemma \ref{lem:general-generator}, we required
that $\epsilon \le 1/\sqrt{D}$. It is reasonable to expect that
$\delta/Jn < 1$ ( meaning that the desired error is at most the
maximum energy in the system) which implies that this condition is
met. The algorithm has $n$ iterations in which $\bigO(N^2)$ possible
extensions for the MPS are considered. For each such possibility, we
perform a contraction of tensors $(\Bl, \BB, \BB')$ in order to
evaluate the energy of a particular term. This contraction takes
time $\bigO(D^3 d^2)$. Thus the total running time is $\bigO(n N^2
D^3 d^2)$.

%
% ======================================================================
%

\section{Commuting Hamiltonian in 1D}
\label{sec:commuting} 

We now prove \Thm{thm:comm}.  Let us first notice that 
\Thm{thm:mps} immediately implies the first claim in \Thm{thm:comm},
namely that approximating the ground state and ground energy of a
commuting Hamiltonian in 1D to within polynomially good accuracy can
be done efficiently.  This follows from the well known fact that the
ground state of a commuting Hamiltonian in 1D can be described by an
MPS of constant bond dimension.  We can therefore apply
\Thm{thm:mps} to the problem, and hence approximate both the ground
state and ground energy efficiently. 

For completeness, here is a sketch of a proof of this fact: assume
we have a $2$-local commuting Hamiltonian in 1D. If the Hamiltonian
is $k$-local for $k>2$, just combine adjacent particles together.
To see that there is a ground state which is described by an MPS of
constant bond dimension, notice that for any commuting Hamiltonian,
there is a ground state $\ket{\psi}$ which is an eigenvector of each
of the terms in the Hamiltonian, with some well defined eigenvalue
for each term. For each term, consider the projection onto the
eigenspace corresponding to that eigenvalue. For any state with
non-zero projection on the ground state, applying these projections
(no matter the order) would result in a ground state. Since there is
always a computational basis state $\ket{w}$ that has a non-zero
projection on the ground state, we can express a ground state as the
projection of all these local terms applied to $\ket{w}$.  We first
apply the projections which interact the pairs of particles $(1,2)$,
$(3,4)$, etc; we then apply the projections that interact the pairs
of particles $(2,3)$, $(4,5)$, etc.  This sequence of operations can
be viewed as a tensor network of depth $2$.  We can thus represent
the ground state as the contraction of a tensor network of depth
$2$. It can be easily seen that such a state must have a constant
Schmidt rank along any cut between the left and right sides; to move
to an MPS of a constant bond dimension, use Vidal's result
\cite{vidal1}. 

Let us now provide the proof of the improvement to an exact
algorithm, for the case that the Hamiltonian has
a polynomial spectral gap. In other words, we
are promised that the ground energy is separated from the rest of
the eigenvalues of the Hamiltonian by a gap $\Delta\ge 1/n^c$ for
some constant $c$. Notice that we don't assume a unique ground state.

The first step of the proof would be to use \Thm{thm:mps} to find
an MPS $\ket{\Omega}$ of constant bond dimension such that
$\bra{\Omega}H\ket{\Omega} \le E_0 + \Delta/3$. From the discussion
above, it is clear that this can be done in polynomial time. Next,
we would like to project this MPS sequentially on some chosen
eigenspaces of the Hamiltonians along the chain. As we are in a
commuting system, this would result in a common eigenvector of all
Hamiltonians, and therefore an eigenvector of $H$ itself. If we
manage to do this without increasing the energy above $E_0 +
\Delta$, then by the existence of the gap, we are promised to have
reached a ground state.

To do this, we rely on the following lemma:
\begin{lemma}
  Let $H=\sum_i H_i$ be a commuting local Hamiltonian system with
  ground energy $E_0$, and let $\ket{\psi}$ be a state such that
  $\bra{\psi}H\ket{\psi} = E_0 + h$. Consider one term $H_i$ in $H$
  with $k$ eigenvalues and projections $P_1, \ldots, P_k$ into the
  corresponding eigenspaces. For every $j=1,\ldots, k$, let
  $\ket{\psi_j}$ be the \emph{normalization} of $P_j\ket{\psi}$, and
  let $c_j = \bra{\psi}P_j\ket{\psi}$. Then for any $n>2$ 
  there is always a $j$ such that $c_j\ge
  \frac{1}{kn^2}$ and $\bra{\psi_j}H\ket{\psi_j} \le E_0 +
  (1+\frac{1}{n})h$.
\end{lemma}
\begin{proof}
  As the $\{H_i\}$ terms are commuting, it follows that
  \begin{align*}
    \bra{\psi}H\ket{\psi} &= \bra{\psi}P_1 H P_1\ket{\psi} +
        \bra{\psi}P_2 H P_2\ket{\psi} \\
        &\ \ +  \ldots + 
        \bra{\psi}P_k H P_k\ket{\psi} \\
      &= c_1 \bra{\psi_1}H\ket{\psi_1} +
        c_2 \bra{\psi_2}H\ket{\psi_2} \\
        &\ \ +  \ldots +
        c_k \bra{\psi_k}H\ket{\psi_k} \ ,
  \end{align*}
  with $\sum_{j=1}^k c_j = 1$. Assume, by contradiction, that for
  every $j$, either $c_j<\frac{1}{kn^2}$ or
  $\bra{\psi_j}H\ket{\psi_j} > E_0 + (1+\frac{1}{n})h$.  Then
  partition the $k$ eigenspaces into two subsets: subset $A$ in
  which the first condition holds, and subset $B$ in which the
  second condition holds. Then
  \begin{align*}
    E_0 + h &= \bra{\psi}H\ket{\psi}\\ 
      &= \sum_A c_j \bra{\psi_j}H\ket{\psi_j}
      +\sum_B c_j \bra{\psi_j}H\ket{\psi_j} \\
       &\ge E_0\sum_A c_j 
      + \left( E_0 +(1+\frac{1}{n})h\right)
      \sum_B c_j \\
      &= E_0+(1+\frac{1}{n})h\sum_B c_j\ ,
\end{align*}
using $\sum_j c_j=1$. 
Since $\sum_A c_j \le \frac{k}{kn^2}=\frac{1}{n^2}$, we have that 
$\sum_B c_j \ge 1-\frac{1}{n^2}$. 
Plugging this into the above equality implies 
$h> h(1+\frac{1}{n})(1-\frac{1}{n^2})$ which is a contradiction for
$n>2$. 
\end{proof}  

We now apply the lemma sequentially to project the approximate state
$\ket{\Omega}$ on the relevant local eigenspaces. We start with
$H_{1,2}$, where we use $h=\Delta/3$ in the lemma. 
The lemma promises the existence of a subspace indexed $j$ (out of $k$
possible $j$s) which, if we project $\ket{\Omega}$ onto that
subspace, the projection will not have too large energy. We denote $c_j$
and $P_j$ by $c_{12}$ and $P_{12}$ respectively (We will shortly
explain how all calculations required for finding the promised $j$ 
can be done efficiently). We
proceed to find $c_{23}$ and $P_{23}$ for the next term $H_{2,3}$,
using the newly projected state, 
and so on up to $H_{n,n-1}$.  After
applying the $n-1$ projections, using the lemma $n-1$ times, we arrive to a
 state $\ket{\psi}$ given by
\begin{align*}
  \ket{\psi} = \frac{1}{\sqrt{c_{12}c_{23}\cdots c_{n-1,n}}}
    P_{12}P_{23}\cdots P_{n-1,n} \ket{\Omega} \ ,
\end{align*}
which satisfies
\begin{align*}
  \bra{\psi}H\ket{\psi} \le E_0 +
    \left(1+\frac{1}{n}\right)^{n-1} \frac{\Delta}{3} \le E_0+
    \frac{e\Delta}{3} \ .
\end{align*}
Using the assumption of the gap and the fact that $\ket{\psi}$ is an
eigenvector of $H$, it must be that $\ket{\psi}$ is a ground state
and $\bra{\psi}H\ket{\psi}=E_0$.

We now argue why finding the $j$ whose existence is promised by the
lemma can be done efficiently. Consider for example the term
$H_{m,m+1}$. To find the relevant $j$ we have to compute, for the
current state $\ket{\psi}$, both the norms squared
$c_j=\bra{\psi}P_j\ket{\psi}$ as well as the expectation values
$\bra{\psi_j}H\ket{\psi_j}=\frac{1}{c_j}\bra{\psi}P_j H
P_j\ket{\psi}$, for all eigenspaces $P_j$ of $H_{m,m+1}$.  Note
first that we are handling here real numbers; the projections $P_j$
on the eigenspaces of $H_{m,m+1}$ may require
infinite precision to describe exactly in binary (or any other)
representation.  We truncate the entries in the projections to
exponentially good precision, using polynomially many bits, so that
all the calculations can be performed efficiently.  This introduces
an exponentially small error.  

\begin{figure}
  \begin{center}
    \includegraphics[scale=0.72]{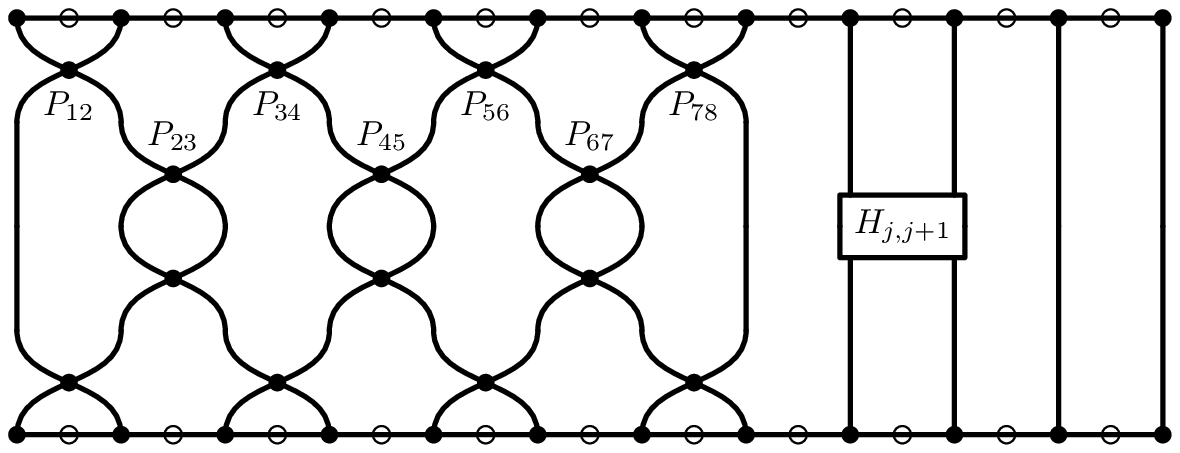}
  \end{center}
  \caption{An illustration of how the expression in \Eq{eq:tnet} is
    given as a tensor-network with a constant number of horizontal
    layers. Specifically, the figure describes the tensor-network of
    $\langle \Omega | P_{12}P_{23}\cdots\, H_{j,j+1}\, \cdots
    P_{23}P_{12}|\Omega\rangle$  
  \label{fig:tnet} }
\end{figure}

The expressions we are interested in calculating are all of the form
\begin{align}
\label{eq:tnet}
  \bra{\Omega}P_{12} \cdots P_{m,m-1}\cdot P_j O
    P_j \cdot P_{m,m-1} \cdots P_{12}\ket{\Omega} \ ,
\end{align}
where $O$ can be either a local Hamiltonian $H_{i,i+1}$ or the identity,
and the $P_{i,i+1}$ are projections on eigenspaces of
the local terms.  Recalling that $\ket{\Omega}$ is a constant-bond
MPS, and using the fact that the projections commute between
themselves, we can write \Eq{eq:tnet} as a constant depth-tensor
network. This is done by partitioning the projections into two
layers: in one layer the projections that work on the sites $(1,2),
(3,4), (5,6), \ldots$ and on the other, the projections that act on
the sites $(2,3), (4,5), (6,7), \ldots$. The resultant
tensor-network is shown in \Fig{fig:tnet}. One dimensional
tensor-networks with constant depth can be efficiently calculated on
a classical computer because their bubble width is constant when
swallowed from left to right
\cite{aharonov2006quantum}.\footnote{This can also be seen by
analyzing the tree-width of that tensor-network and using the
analysis of \Ref{markov2008simulating}}

Thus, all calculations (under our assumptions of polynomially many
bits of precision of the $P_j$'s) can be performed efficiently.  The
resulting state is given by a tensor network of constant depth
(namely the original $\ket{\Omega}$ on which the chosen projections
are applied.) As before, this can be modified to a MPS of constant
bond dimension using Vidal's result \cite{vidal1}, concluding the proof.

{\hfill\qed}

%
%-----------------------------------------------------------------------
%

\subsection{A Proof for the Commuting Hamiltonians case, based on
Ref.~\cite{bravyi}}

First we describe the alternate algorithm assuming we have the
ability to perform arithmetic operations with infinite
precision and then discuss the consequences of limited precision.
\Ref{bravyi} prove certain properties about the ground states of
$2$-local commuting Hamiltonians in which the interaction graph is a
general graph. We express those properties for the special case of
interest here in which the graph is a line. Let $\calh_j$ be the
Hilbert space of particle $j$. It is shown in \Ref{bravyi} that
when the terms of the Hamiltonian commute, the Hilbert space of
each particle can be expressed as a direct sum, $\calh_j =
\oplus_{\alpha_j} \calh_j^{(\alpha_j)}$, such that each
$\calh_j^{(\alpha_j)}$ can then be expressed as a tensor product of 
three spaces
\begin{align*}
  \calh_j^{ (\alpha_j) } = \calh_{L,j}^{ (\alpha_j)} 
    \otimes \calh_{C,j}^{(\alpha_j)}  
    \otimes \calh_{R,j}^{ (\alpha_j)} \ .  
\end{align*}
This structure has the property that $H_{j,j+1}$ leaves the
subspaces $ \calh_j^{ (\alpha_j)} \otimes
\calh_{j+1}^{(\alpha_{j+1})}$ invariant, and moreover, when
restricted to such a subspace, $H_{j,j+1}$ acts non-trivially only
on $ \calh_{R,j}^{ (\alpha_j)} \otimes
\calh_{L,j+1}^{(\alpha_{j+1})}$ (the right part of $\calh_j^{
(\alpha_j)}$ and the left part of $\calh_{j+1}^{ (\alpha_{j+1})}$).
Consequently, there exists a ground state which resides in some
subspace $\calh^{(\alpha)} = \otimes_j \calh_j^{ (\alpha_j)}$, for
some choice of $\alpha_1,\ldots,\alpha_n$.  Moreover, within the
subspace $\calh^{(\alpha)}$ the state can be written as a tensor
product of $2$ particle states living in the spaces of the form $
\calh_{R,j}^{ (\alpha_j)} \otimes \calh_{L,j+1}^{(\alpha_{j+1})}$,
tensored with some arbitrary single particle states living in the
$\calh_{C,j}^{ (\alpha_j)}$ spaces.

If the algorithm knows the correct choice of indices
$\alpha_1,\ldots,\alpha_n$, it can find such a ground state
efficiently, as follows. Note that the descriptions of both the
spaces $\calh_j^{ (\alpha_j)}$ and their divisions $\calh_j^{
(\alpha_j)} = \calh_{L,j}^{ (\alpha_j)} \otimes \calh_{C,j}^{
(\alpha_j)} \otimes \calh_{R,j}^{ (\alpha_j)}$ are derived from
local properties of $\calh_j$ imposed by the two local Hamiltonians
$H_{j-1,j}$ and $H_{j,j+1}$. The subdivision of $\calh_j$ in this
way can be expressed as a solution to a set of quadratic homogeneous
constraints. Since the dimension of $\calh_j$ and hence the number
of variables is constant, it can be efficiently computed. If the
algorithm knows the $\alpha_j$'s, it therefore knows the description
of the subspaces $ \calh_{L,j}^{ (\alpha_j)} \otimes
\calh_{C,j}^{(\alpha_j)} \otimes \calh_{R,j}^{ (\alpha_j)}$, and the
restriction of the $H_{j,j+1}$ to those spaces; it therefore just
needs to find a ground state of linearly many 2-particle
Hamiltonians, which is an easy task. It is therefore enough for the
algorithm to find the correct $\alpha_1,\ldots,\alpha_n$ indices.

We will do this using dynamic programming.  The critical point in
using dynamic programming here is that the energy contribution of
$H_{\ell,\ell+1}$ depends only on the choice of $\alpha_\ell$ and
$\alpha_{\ell+1}$, so the choice of $\alpha_{k}$ for $k \le j-1$
does not affect the energy of the $H_{\ell,\ell+1}$ terms  for any $\ell
\ge j$. Using this observation, the algorithm proceeds from left to
right as follows. For the first term $H_{1,2}$, the algorithm finds
the division into a direct sum of subspaces for particles $1$ and
$2$. The algorithm keeps an optimal state (choice of $\alpha_1$) and
energy for each possible $\alpha_2$.

Then, in a general step, we assume at particle $i$ we have the
following information for each index $\alpha_i$: a list of indices
$\alpha_1,\ldots,\alpha_{i-1}$ such that the ground energy of the
Hamiltonian of particles $1,\ldots,i$ restricted to the subspaces
$\calh_1^{ (\alpha_1)}\otimes\ldots\otimes \calh_i^{ (\alpha_i)}$ is
minimal.  To continue to the next particle, we first compute the
division into subspaces for particle $i+1$, indexed by
$\alpha_{i+1}$, and optimize for each subspace in turn. For each
subspace, we consider all items in the previous list; for each item,
we have a list of subspaces, one for each particle.  We compute the
minimal energy for each such restriction, including now the
$H_{i,i+1}$ term in the calculation of the energy, restricted
according to subspaces $\alpha_{i+1}$ and $\alpha_{i}$, the last
choice coming from the list.  We pick the tail of the subspace of
the $i+1$ particle to be the one which minimizes the terms up to
that point.

Notice that in each step the dynamic program compares partial
energies emerging from restricting the state to a different sector
in the Hilbert space. These energies can be computed efficiently with
polynomially many bits, namely up to exponentially good precision.
Thus, this second algorithm achieves exact results for a somewhat
larger set of Hamiltonians than our first algorithm, namely those
for which 
the partial energies will  not be confused if the computations are
done with exponentially good precision.  

Note that even with
this extremely good resolution, it might be the case that the ground
energy is confused with a slightly excited energy which is, say,
\emph{doubly} exponentially close. We do not know of a good
condition which would rule out the possibility of such very close
energies, except for some very trivial assumptions such as requiring
that all eigenvalues are integer numbers. For example,
even if we require that the different entries in the terms in the
Hamiltonian are all rationals smaller than $1$ with denominator
upper bounded by a constant, it is still not known how to rule out
the possibility that two eigenvalues of the overall Hamiltonian are
doubly exponentially close. This issue touches upon an open question
in number theory related to sums of algebraic numbers -- see the
open problem described in \Ref{p33}, which can be traced back to
\Ref{or} (if not earlier), and also \Ref{qian} and references
therein.

%%%%%%%%%%%%%%%%%%%%%%%%%%%%%%%%%%%%%%%%%%%%%%%%%%%%%%%%%%%%%%%%%%%%%%%%%%%

\section{Acknowledgment}
\label{sec:acknowledgment}

Dorit Aharonov is partially supported by ISF Grants 039-7549 and
039-8066, ARO Grant 030-7799, and SCALA Grant 030-7811. Sandy Irani
Partially supported by NSF Grant CCF-0916181. Itai Arad acknowledges
support by the ERC Starting grant of Julia Kempe (PI).

The main progress on the results reported on in this paper was made
while the three authors were visiting the Erwin Schr\"{o}dinger
International Institute for mathematical Physics (ESI) in Vienna,
Austria.

%%%%%%%%%%%%%%%%%%%%%%%%%%%%%%%%%%%%%%%%%%%%%%%%%%%%%%%%%%%%%%%%%%%%%%%%%%%
%%%%%%%%%%%%%%%%%%%%%%%%%%%%%%%%%%%%%%%%%%%%%%%%%%%%%%%%%%%%%%%%%%%%%%%%%%%
%%%%%%%%%%%%%%%%%%%%%%%%%%%%%%%%%%%%%%%%%%%%%%%%%%%%%%%%%%%%%%%%%%%%%%%%%%%
%%%%%%%%%%%%%%%%%%%%%%%%%%%%%%%%%%%%%%%%%%%%%%%%%%%%%%%%%%%%%%%%%%%%%%%%%%%
%%%%%%%%%%%%%%%%%%%%%%%%%%%%%%%%%%%%%%%%%%%%%%%%%%%%%%%%%%%%%%%%%%%%%%%%%%%
%%%%%%%%%%%%%%%%%%%%%%%%%%%%%%%%%%%%%%%%%%%%%%%%%%%%%%%%%%%%%%%%%%%%%%%%%%%

\appendix*

%%%%%%%%%%%%%%%%%%%%%%%%%%%%%%%%%%%%%%%%%%%%%%%%%%%%%%%%%%%%%%%%%%%%%%%%%%%

\section{Proofs of lemmas}

\label{sec:proofs}

\noindent\emph{Proof of Lemma \ref{lem:general-generator}:}

Let $\delta = \nu/72 b$. We will occasionally use the
assumption that $\delta \le 1/72 b$.

First we create a set $R(\delta)$ of real numbers in the interval
$[0,1]$ such that for any real number in the range $[0,1]$, it is
within $\delta$ of some element in $R(\delta)$. $R(\delta)$ will
have $\lceil 1 / 2\delta \rceil$ elements. To create $R(\delta)$, we
add $(2j+1)\delta$ for each integer j in the range from $0$ through
$\lceil 1 / 2\delta \rceil-2$. Note that the largest point in
$R(\delta)$ so far is in the range $[1-3\delta, 1 -\delta)$. Then we
add $1 - \delta$ to $R(\delta)$.

Then using $R(\delta)$, we create a set $C(\delta)$ which is a set
of complex scalars which form a net over all complex scalars with
norm at most $1$. Include $ x e^{i 2 \pi y} $, for every $x,y \in
R(\delta)$. There are $\lceil 1/2 \delta\rceil^2 \le (1/\delta)^2$
points in $C(\delta)$. For any complex number $c$ if norm at most
$1$, there is a number $c'$ in $C(\delta)$ such that $|c-c'| \le 2
\delta$.

To generate $S_{a,b}$, consider first the set $S_1$ of of all
possible $a\times b$ matrices with entries from $C(\delta)$. This
set contains $|C(\delta)|^{ab}$ matrices.  In the case where $a=1$
and we only want entries with real, non-negative coefficients, we
use $R(\delta)$ for the entries instead of $C(\delta)$ and the set
contains $|R(\delta)|^{b}$ matrices (in fact, vectors).  Then: 
\begin{enumerate}
  \item
    Remove any matrix from $S_1$ which has a row whose norm is 
    greater than $1+\sqrt{b} 2 \delta$ or less than $1 - \sqrt{b} 2
    \delta$, to get $S_2$.
  
  \item Renormalize each row in every matrix in $S_2$ to get $S_3$.

  \item Remove any matrix from $S_3$ which has any two rows whose inner product
    is more than $9 \sqrt{b} \delta$.
  
  \item For any matrix in $S_3$, Apply the Gram-Schmidt
    procedure to the rows to form an orthonormal set.  
\end{enumerate}

We claim that the final set is the desired $S_{a,b}$. 
Note that the number of matrices is $O((1/\delta)^{2ab})=O((72 b/\nu)^{2ab})$, 
and the running time to produce the set is 
$O(a^2 b (1/\delta)^{2ab} )=O(a^2 b (72 b/\nu)^{2ab} )$ as required. 
What remains to show is that if
$A$ is any $a \times b$ matrix whose rows form an ortho-normal set then
we can find a matrix $B$ in $S_{a,b}$ which is close to it. 

Let $W$ be an $a \times b$ matrix. We will denote it's $i^{th}$ row
by $W_i$. Define the distance between two matrices $d(W,W')$ to
be $\max_i \| W_i - W'_i \|$.
Let $X$ be the matrix obtained by rounding every entry in $A$ to
the nearest complex number in $C(\delta)$. Let $Y$ be the matrix 
obtained after the
rows of $X$ are normalized and let $Z$ be the matrix obtained after the
rows of $Y$ are transformed into an 
ortho-normal set via the Gram-Schmidt procedure.
We need to prove that $d(A,Z)\le \nu$, and to show that $Z\in S_{a,b}$, 
which would imply together that we can choose $B$ in the lemma 
to be equal to $Z$.   

We will now prove both of the above claims. 
For the second part we need to  
prove that $X$ survives step $1$ and $Y$ survives step $3$.

{~}

\noindent
{\bf $X$ survives step $1$:}
Since each entry in $A-X$ has magnitude at most $2\delta$,
we know that $d(A,X) \le \sqrt{b} 2 \delta$.
In order to bound the norm of $X_i$, observe that 
$$ \sqrt{b} 2 \delta \ge \norm{ A_i - X_i} \ge | \norm{ A_i} - \norm{X_i} |.$$
Since $\norm{A_i}=1$, it follows that
 $\norm{X_i}$ lies in the range from $1 -\sqrt{b}2\delta$ to
$1 + \sqrt{b}2 \delta$ and it will survive Step $1$.
We have: 
\begin{align*}
  d(X,Y) &\le \max_i \| A^i - \frac{1}{1 -  \sqrt{b}2 \delta} A^i \| \\
  &= \frac{\sqrt{b}2 \delta}
    {1 -  \sqrt{b}2 \delta} \le \sqrt{b}2 \delta(36/35) \ .
\end{align*}
The latter inequality uses the assumption that
$\delta \le 1/72 \sqrt{b}$.
Using the triangle inequality for our distance $d(\cdot)$,
we have that 
for any $i$ $\norm{A_i - Y_i} \le (4 + \frac{2}{35}) \sqrt{b} \delta$.

{~}
\noindent
{\bf $Y$ survives step $3$:} 
Now we need to bound the inner product of any two rows of $Y$ in
order to establish that it is not removed in Step $3$: 
\begin{align*}
| \braket{Y_i}{Y_j}| 
  &= |\braket{A_i + (Y_i-A_i)  }{ A_j + (Y_j-A_j)}|\\
 &\le | \braket{A_i}{A_j} | 
     + | \braket{ Y_i-A_i }{ Y_j-A_j } | \\
 &\ \ + | \braket{ Y_i-A_i }{ A_j  } |
     + | \braket{ A_i  }{ Y_j-A_j } | \\
 &\le 
 \norm{ Y_i-A_i }\norm{ Y_j-A_j }  +
 \norm{ Y_i-A_i }\norm{ A_j  } \\
 &\ \ + \norm{ A_i  }\norm{ Y_j-A_j }  \\
 &\le \sqrt{b} \delta \left[ 
     \left( 4 + \frac{2}{35} \right)^2 \sqrt{b} \delta
    + 2 \left( 4 + \frac{2}{35}  \right)
   \right] \\
  &\le 9 \sqrt{b} \delta \ .
\end{align*}

The second inequality uses the Chauchy-Schwartz inequality.
The last inequality uses the fact that $\sqrt{b} \delta \le 1/72$.

{~}
{\bf Bounding the distance $d(A,Z)$:} 
Finally, we need to consider how much the matrix shifts as a result
of the Gram-Schmidt procedure, to bound $d(Y,Z)$.
Let $\mu = 9 \sqrt{b} \delta= 9 \nu / 72 \sqrt{b}$.
Since $a \le b$, by assumption in the lemma, we know that
$\mu \le 9 \nu/ 72 \sqrt{a}$.
We use this latter bound in the next part of the proof since we
are bounding quantities by a function of $a$ instead of $b$.
Since we assume that $\nu \le 1/\sqrt{a}$, we can
assume that $a \mu \le 9/72$.
Recall that the Gram-Schmidt procedure starts with $Z_1 = Y_1$.
Then each $Z_i$ is determined by first creating an unnormalized
state $\tilde{Z}_i$:
$$\tilde{Z}_i = Y_i - \sum_{j=1}^{i-1} \braket{Z_j}{Y_i} Z_j.$$
Then $\tilde{Z}_i$ is normalized to $1$.
We will prove the following two properties by induction in $i$, 
\begin{enumerate}
\item $| \braket{Z_i}{Y_j} | \le 2\mu$
for all $j$ such that  $j > i$ 
\item $1 -  2 \sqrt{a} \mu \le \norm{ \tilde{Z}_i } \le 1 +  2 \sqrt{a} \mu$.
\end{enumerate}
$\tilde{Z}_1$ is
not defined, but we can take it to be $Z_1$.
The two properties clearly hold for $Z_1$. 
Now by induction
\begin{align*}
\norm{\tilde{Z}_i } = & \norm{ Y_i - \sum_{j=1}^{i-1} \braket{Z_j}{Y_i} Z_j}\\
\le & \norm{Y_i} + \norm{ \sum_{j=1}^{i-1} \braket{Z_j}{Y_i} Z_j }\\
= & 1 +   \left( \sum_{j=1}^{i-1}  \braket{Z_j}{Y_i} \braket{Y_i}{Z_j} \right)^{1/2} \\
\le & 1 +  2 \sqrt{a} \mu
\end{align*}

A similar argument can be used to show that $1 - 2 \sqrt{a} \mu \le \norm{\tilde{Z_i}}$.
Next we establish Property $1$:
\begin{align*}
| \braket{ Y_k }{ Z_i } | = &
\frac{1}{\norm{\tilde{Z}_i }} \left| \braket{Y_k}{Y_i}
- \sum_{j=1}^i \braket{Z_j}{Y_i} \braket{Y_k}{Z_j} \right| \\
\le & \frac{1}{1 - 2 \sqrt{a} \mu}  \left[ \mu
+ \sum_{j=1}^{i-1} 4 \mu^2 \right]  \\
\le & \frac{ \mu( 1 + 4 a \mu)}{1 - 2 \sqrt{a} \mu} \le 2 \mu
\end{align*}
The first inequality follows from the inductive hypothesis.
The last inequality make use of the fact that $a \mu \le 9/72$.
Finally to bound $\norm{Y_i - Z_i}$, we have
\begin{align*}
| \norm{ Y_i - Z_i } | \le &
\left(1- \frac{1}{\norm{\tilde{Z}_i }}\right) \norm{Y_i} 
+ \frac{1}{\norm{\tilde{Z}_i }}\norm{ \sum_{j=1}^{i-1} \braket{Z_j}{Y_i} Z_j} \\
\le & \frac{ 2 \sqrt{a} \mu \norm{Y_i} }{ 1 - 2 \sqrt{a} \mu } + \frac{1}{1 - 2 \sqrt{a} \mu}  \left(  \sum_{j=1}^{i-1} |\braket{Z_j}{Y_i}|^2 \right)^{1/2}\\
\le & \frac{ 4 \sqrt{a} \mu  }{ 1 - 2 \sqrt{a} \mu } \le 6 \sqrt{a} \mu  
\end{align*}

The last inequality uses again the fact that $\sqrt{a} \mu \le 9/72$.
The total distance between $A$ and $Z$ is at most 
$5 \sqrt{b} \delta + 6 \sqrt{a} \mu $. Plugging
in $\mu = 9 \sqrt{b} \delta  $ and using the fact that 
$a \le b$, we get an upper
bound of $59 b \delta\le \nu$ on the distance of $A$ to $Z$, 
using the definition of $\delta$. 
{\hfill\qed}

\begin{lemma}
\label{lem:Schmidt}
  Let $\ket{A}$, $\ket{B}$ be two two-particles states
  that, and expand them in the standard basis of the first particle:
  \begin{align*}
    \ket{A} &= \sum_i a_i\ket{i}\ket{A_i} \ , \\
    \ket{B} &= \sum_i b_i\ket{i}\ket{B_i} \ ,
  \end{align*}
  such that $\ket{A_i}$ are normalized but not-necessarily orthogonal
  to themselves and similarly the $\ket{B_i}$.
  Then
  \begin{align}
    \norm{a-b} = \left( \sum_i |a_i-b_i|^2\right)^{1/2} \le \norm{A-B} \ .
  \end{align}
\end{lemma}

\begin{proof}

  \begin{align*}
    \norm{a-b}^2 &=  \sum_i |a_i-b_i|^2 \\
      &\le \sum_i \norm{a_i\ket{A_i} - b_i\ket{B_i}}^2 \\
      & = \norm{\sum_i \ket{i}(a_i\ket{A_i} - b_i\ket{B_i})}^2 \\
      & = \norm{\ket{A} - \ket{B}}^2 \ .
  \end{align*}  
  
\end{proof}

\bibliography{onedimbib}

\end{document}